\documentclass[a4paper,fleqn]{cas-dc}

\usepackage[authoryear,longnamesfirst]{natbib}
\usepackage[most]{tcolorbox} 
\usepackage{amsmath}
\def\tsc#1{\csdef{#1}{\textsc{\lowercase{#1}}\xspace}}
\tsc{WGM}
\tsc{QE}
\begin{document}
\let\WriteBookmarks\relax
\def\floatpagepagefraction{1}
\def\textpagefraction{.001}

% Short title
\shorttitle{}    

% Short author
\shortauthors{}  

% Main title of the paper
\title [mode = title]{DALI LLM-Agent Enhanced Dual-Stream Adaptive Leadership Identification for Group Recommendations}  

% Title footnote mark
% eg: \tnotemark[1]
\tnotemark[1] 

% Title footnote 1.
% eg: \tnotetext[1]{Title footnote text}
\tnotetext[1]{} 

% First author
%
% Options: Use if required
% eg: \author[1,3]{Author Name}[type=editor,
%       style=chinese,
%       auid=000,
%       bioid=1,
%       prefix=Sir,
%       orcid=0000-0000-0000-0000,
%       facebook=<facebook id>,
%       twitter=<twitter id>,
%       linkedin=<linkedin id>,
%       gplus=<gplus id>]

\author[1]{Boxun Song}%[<options>]

% Corresponding author indication
% \cormark[1]

% Footnote of the first author
\fnmark[1]

% Email id of the first author
\ead{202424021053t@stu.cqu.edu.cn}

% URL of the first author
\ead[url]{}

% Credit authorship
% eg: \credit{Conceptualization of this study, Methodology, Software}
\credit{Conceptualization, Data curation, Formal analysis, Methodology, Software Validation, Writing - original, Writing - review & editing}

% Address/affiliation
\affiliation[1]{organization={School of Big Data and Software Engineering},
            addressline={Chongqing University}, 
            city={Chongqing},
%          citysep={}, % Uncomment if no comma needed between city and postcode
            postcode={401331}, 
            state={Chongqing},
            country={China}}

\author[2]{Ming Gao}%[]
 \cormark[1]
\fnmark[2]
\ead{}
\ead[url]{}
\credit{Funding acquisition, Resources, Supervision, Writing - review & editing}
\affiliation[2]{organization={School of Big Data and Software Engineering},
            addressline={Chongqing University}, 
            city={Chongqing},
%          citysep={}, % Uncomment if no comma needed between city and postcode
            postcode={401331}, 
            state={Chongqing},
            country={China}}

\author[3]{Jiawei Chen}%[]
%  \cormark[1]
\fnmark[3]
\ead{}
\ead[url]{}
\credit{Writing - review & editing}
\affiliation[2]{organization={School of Big Data and Software Engineering},
            addressline={Chongqing University}, 
            city={Chongqing},
%          citysep={}, % Uncomment if no comma needed between city and postcode
            postcode={401331}, 
            state={Chongqing},
            country={China}}

% Corresponding author text
\cortext[2]{Corresponding author}

% Footnote text
\fntext[1]{}

% For a title note without a number/mark
%\nonumnote{}

% Here goes the abstract
\begin{abstract}
Group recommendation systems play a pivotal role in supporting collective decisions across various contexts, from leisure activities to organizational team-building. Existing group recommendation approaches typically use either handcrafted aggregation rules (e.g. mean, least misery, weighted sum) or neural aggregation models (e.g. attention-based deep learning frameworks), yet both fall short in distinguishing leader-dominated from collaborative groups and often misrepresent true group preferences, especially when a single member disproportionately influences group choices. To address these limitations, we propose the Dual-stream Adaptive Leadership Identification (DALI) framework, which uniquely combines the symbolic reasoning capabilities of Large Language Models (LLMs) with neural network-based representation learning. Specifically, DALI introduces two key innovations: a dynamic rule generation module that autonomously formulates and evolves identification rules through iterative performance feedback, and a neuro-symbolic aggregation mechanism that concurrently employs symbolic reasoning to robustly recognize leadership groups and attention-based neural aggregation to accurately model collaborative group dynamics. Experiments conducted on the Mafengwo travel dataset confirm that DALI significantly improves recommendation accuracy compared to existing frameworks, highlighting its capability to dynamically adapt to complex, real-world group decision environments.
\end{abstract}

% Use if graphical abstract is present
%\begin{graphicalabstract}
%\includegraphics{}
%\end{graphicalabstract}

% Research highlights
\begin{highlights}
\item %\textbf{Pioneering the use of LLM-powered agents for dynamic leadership identification in group recommendations.} This work is the first framework, to the authors' knowledge, that employs LLM-based agents to autonomously identify groups with dominant leaders. It transitions the paradigm from static, heuristic rules to an adaptive, evidence-driven process where agents continuously generate and refine identification criteria through iterative performance feedback.
\item Proposes the first LLM-agent-driven framework for dynamic leadership identification in groups.
\item Designs a self-optimizing closed-loop rule engine for “practice-reflection-evolution”.
% \textbf{Design of a self-optimizing, closed-loop rule evolution engine.} A core innovation is the development of a dynamic rule-creation engine. This engine establishes a "practice-reflection-evolution" closed loop, enabling LLM agents to monitor shifts in member influence and translate these patterns into updatable, quantifiable decision rules. This mechanism allows the framework to dynamically adapt to evolving group power structures, overcoming the rigidity of static rule-based methods.
\item Constructs a synergistic neuro-symbolic dual-stream hybrid decision architecture.
%\textbf{Proposal of a synergistic neuro-symbolic hybrid aggregation mechanism.} The framework introduces a novel dual-stream architecture that synergistically integrates the interpretability of symbolic reasoning with the representation learning capability of neural networks. It employs a symbolic channel for rule-based classification and a parallel neural channel for capturing implicit patterns, with a triple-threshold decision logic for integration. Ablation studies confirm that the full hybrid framework significantly outperforms variants using either stream in isolation.
\end{highlights}

%\nocite{*}

% Keywords
% Each keyword is seperated by \sep

\maketitle

% Main text

\section{Introduction}\label{}

Nowadays, computer systems store vast amounts of personal data. \cite{nguyen2025survey} The emergence of recommendation systems is precisely to address the problem that users find it difficult to retrieve the data they need from the massive data pool.

Group recommendation systems aim to generate recommendations \cite{li2024recent} that satisfy the preferences of the majority within a group. In recent years, collective activities, such as family gatherings and corporate team events, have proliferated, highlighting the growing significance of these systems. For instance, in scenarios like family movie selection or group travel planning, group recommendation systems are essential to facilitate collective decision-making \cite{li2023self}.

% systems \cite{li2024recent} facilitate collective decision-making for multi-user activities ranging from family gatherings to corporate events. These systems face the core challenge of modeling collective preferences despite sparse explicit group interactions \cite{li2023self}. 

Group recommendation systems require modeling multi-user collective preferences despite the inherent sparsity of explicit group interactions \cite{cao2019social}. Early approaches relied on predefined rules \cite{malecek2021fairness, sato2022enumerating} for aggregating individual preferences, failing to adequately address real-world complexities arising from conflicting member preferences and dynamic power structures. Subsequent research shifted towards learned aggregation paradigms: AGREE \cite{cao2018attentive} introduced attention mechanisms to adaptively weight member contributions. Similarly, CubeRec \cite{chen2022thinking} modeled groups using hypercubes with distance metrics. Consrec \cite{wu2023consrec} further models the connections between members and groups through multi-perspective learning. LARGE \cite{gan2025large} revealed the existence of leadership structures where influential members dominate decisions, yet relied on static or predefined identification rules rather than learning them dynamically. This progression highlights a critical gap: while learned methods optimize for consensus, they lack explicit modeling of power asymmetries, preventing dynamic adaptation to fundamental group characteristics such as dominance hierarchies.

Recent advancements in Large Language Models (LLMs) offer a transformative approach to leadership identification within group recommendations. Conventional methods face limitations in capturing the fluid nature of group dynamics.To address this, we propose Dual-stream Adaptive Leadership Identification (DALI). These agents facilitate adaptive rule evolution: they continuously refine leadership identification criteria through performance feedback, enabling them to capture evolving group interactions and power structures. Crucially, DALI's dual-stream design leverages the agents' multimodal reasoning capabilities. The agents integrate diverse signals—including member interactions, attention distributions, and temporal patterns—into unified leadership assessments, simultaneously generating interpretable decision traces. This evidence-driven reasoning transforms leadership identification from heuristic to adaptive: by correlating emergent behavioral patterns with contextual features, the agents construct adaptive probabilistic rules. Consequently, DALI resolves the rigidity of predefined rules and the opacity of neural methods, establishing a self-optimizing paradigm for adaptive leadership identification.

DALI operates via dual-stream synergy: The symbolic stream dynamically updates rules via LLMs to respond to power shifts; the neural stream learns influence representations to capture collaborative patterns. Closed-loop feedback enables co-evolution, continuously enhancing identification accuracy.

Our contributions can be summarized as follows: 

\begin{itemize}

\item To the best of our knowledge, DALI is the first framework employs LLM-based agents to identify groups with dominant leaders. This work defines clear, measurable signals that distinguish leader-led groups from those where decisions are made more equally. 
    
\item We develop a novel dynamic rule-creation engine where attention networks continuously detect shifts in member influence, and LLM agents instantly translate these shifts into adaptive leadership rules. This seamless fusion of pattern detection and rule synthesis generates evidence-based, updatable decision criteria.
    
\item Our agent-based framework achieves self-optimization through the continuous refinement of its leadership identification rules. This capability enables dynamic adaptation to evolving group dynamics and power imbalances, overcoming the rigidity of static rule-based methods.
\end{itemize}

\begin{figure*}
  \centering
    \includegraphics[width=1\linewidth,height= 7 cm]
    {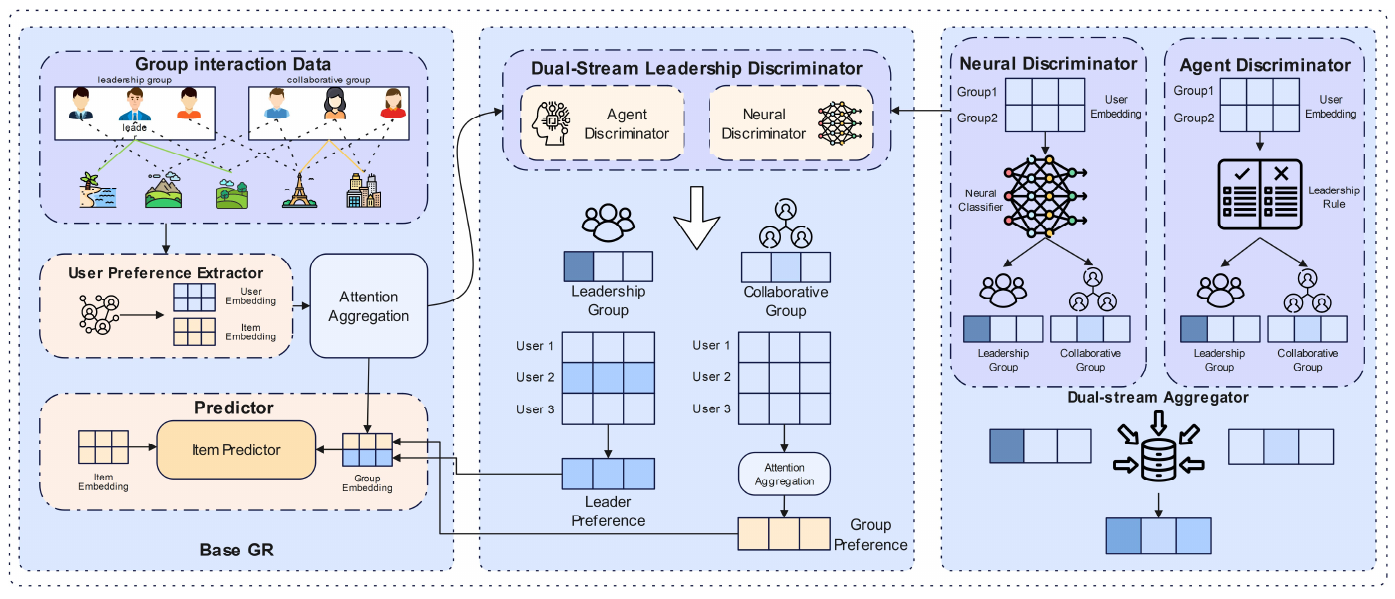}
   \caption{
\textbf{Overall architecture of DALI}. The DALI framework extracts group member weights from interaction data using the base GR model's attention module. These weights are processed by a dual-stream leader group discriminator—comprising neural and symbolic classifiers—to determine group type. The discriminator outputs are aggregated to yield a final classification, which is then fed back into the GR attention module for recommendation.}
   \label{fig:structure}
\label{fig1}
\end{figure*}
\section{Preliminary}\label{}
In this paper, we design a Dual-stream Adaptive Leadership Identification framework (DALI) driven by large language model (LLM)-powered agents. DALI leverages the reasoning capabilities of LLMs to analyze discriminatory rules for identifying leadership groups, thereby enhancing group recommendations. This section first introduces group recommendation modeling, followed by a detailed exposition of DALI’s architecture.

Within recommendation systems, given user-item interaction data $ D=\left\{ u, i, r_{u, i} \mid u \in U, i \in I\right\} $. Let $u \in U$ denote the user and $i \in I$ denote the item. Define a binary interaction indicator $r_{u,i} \in \{0,1\}$. $r_{u,i} = 1$  indicates an observed interaction between user  and item . $r_{u,i} = 0$  denotes no interaction.

Recommendation algorithms process this data to learn latent features, modeling user preferences to derive a user latent representation matrix $\mathbf{Z_{U}} \in |U| \times \mathbf{d}$, an item latent representation matrix $\mathbf{Z_{I}} \in |{I}| \times \mathbf{d}$, and a prediction function $f(\cdot; \theta_{f})$.

For group recommendation systems, the model must further capture a group representation. Let 
$ \mathbf{p_{u}}$ denote the embedding of a group member u. The group representation $ \mathbf{p_{g}}$  is aggregated from member embeddings:
\begin{equation}
\begin{gathered}
    \mathbf{p_{g}} = \sum_{u=1}^{n} \alpha_{u} \cdot\mathbf{p_{u}} .
  \end{gathered}
\end{equation}
Crucially, in leadership groups, group preferences are solely determined by the leader member $u_{0}$. The aggregation thus simplifies to:
\begin{equation}
\begin{gathered}
    \mathbf{p_{g}} = \mathbf{p_{u_{0}}} .
  \end{gathered}
\end{equation}
To accurately identify leadership groups and their leaders, we propose two key innovations. First, an adaptive rule evolution engine triggers LLM-driven rule generation and version control via performance monitoring, establishing a "practice-reflection-evolution" closed loop. Second, the dual-stream adaptive framework (DALI) integrates symbolic rules' interpretability with neural networks' generalization through progressive transition mechanisms for collective decision modeling. 

\section{Methodology}\label{}
As shown in Figure 1, the DALI framework achieves leader group identification through a four-component hierarchical system: employing LLM rule engines for interpretable dynamic weight analysis while concurrently integrating neural networks to learn implicit interaction patterns. Neuro-symbolic hybrid aggregators innovatively fuse dual-channel outputs, ultimately driving end-to-end recommendation generation. The core innovation establishes a closed-loop cognitive system wherein continuous monitoring of recommendation outcomes feeds back into the rule evolution engine, enabling autonomous optimization of leadership discrimination rules via Generative Language Model (GLM) agents. This completes a full cognitive cycle encompassing data input, decision formulation, feedback analysis, and rule evolution—enhancing recommendation performance while consistently maintaining high interpretability of discrimination results.

\subsection{Agent Architecture of DALI}

% 本文框上下间距改一下
% Agent改个名字
% 整个方法把前面的第一段的内容放下去
% 图片排版调整一下
% 公式4拆开
% 公式后面加句号
% 公式3，4的Pg删掉
% introduction加上贡献

% Lgroup写出具体的
% 总损失放在后面，直接叫L
% η改成其他符号
% 实验加一个参数分析和收敛实验

%表1换成大表

% The agent of DALI framework dynamically perceives and optimizes group decision patterns through four synergistic modules: the Role Definition Module establishes dual agent identities (Rule Governance Expert and Evolution Engine); the Updatable Memory Module constructs a dual-channel knowledge architecture with versioned rule repositories and case feature banks; the Planning Module drives self-iteration via a performance-rule feedback loop; and the Action Module executes weight tensor analysis and rule optimization. Modular collaboration achieves the transition from static rules to dynamic cognition, forming a systematic leadership group identification framework. 
Figure 2 illustrates the four synergistic modules that work together in the DALI agent's operation: the Role Module establishes dual agent identities for governance and evolution; the Memory Module constructs a knowledge architecture with versioned rules and case features; the Planning Module drives self-iteration via performance-rule feedback; the Action Module executes decision analysis and rule optimization. This modular collaboration achieves the transition from static rules to dynamic cognition, forming a systematic leadership identification framework.

% \textbf{Role Module}. The Role Module establishes DALI's dual cognitive positioning: the Rule Governance Expert executes dynamic rule governance for real-time leadership/collaborative group classification using weight concentration features; the Rule Evolution Engine drives multi-source fusion rule iteration by integrating historical efficacy, real-time behaviors, and memory insights to generate quantitatively validated rules. 

\textbf{Role Module.} The Role Module establishes the dual cognitive positioning of the DALI framework, comprising two core components: the \textbf{Rule Governance Expert} and the \textbf{Rule Evolution Engine}.

The Rule Governance Expert is responsible for dynamic rule governance. Its core task is to identify leadership patterns within group interactions, primarily by extracting weight distribution features (e.g., weight concentration) and matching them against a repository of discrimination rules, thereby enabling real-time classification and labeling of leadership-oriented versus collaborative-oriented groups.

\begin{tcolorbox}[
    colback=gray!10,    % 背景色（浅灰色）
    colframe=gray!50,   % 边框色（深灰色）
    rounded corners,    % 圆角
    boxrule=0.5pt,      % 边框粗细
    left=3pt, right=3pt, % 左右内边距   
]
As a dynamic rule governance expert, identify leadership patterns by extracting weight distribution features and matching discrimination rules for real-time group classification. 

\end{tcolorbox}

The Rule Evolution Engine is responsible for driving the iterative upgrading of the system's rule repository. It achieves this by integrating multi-source information—including historical rule efficacy data (such as fluctuations in ranking quality metrics like NDCG), real-time group behavioral characteristics (e.g., speed of opinion convergence), and insights from the memory module—to continuously detect obsolete rule entries and propose new, quantifiable rule candidates with computable conditions (e.g., statistics based on weights). This engine follows a version-controlled iterative process to ensure the progressive optimization of the rule system.

\begin{tcolorbox}[
    colback=gray!10,    % 背景色（浅灰色）
    colframe=gray!50,   % 边框色（深灰色）
    rounded corners,    % 圆角
    boxrule=0.5pt,      % 边框粗细
    left=3pt, right=3pt, % 左右内边距   
]

As a rule evolution engine, upgrade the rule system by detecting obsolete rules and proposing quantifiable new rules with computable conditions via version-controlled iteration. 

\end{tcolorbox}

%角色定义模块确立DALI双重认知定位：规则治理专家执行动态规则治理，利用权重集中特征实现领导型/协作型群组实时分类；规则演化引擎驱动多源融合规则迭代，整合历史效能、实时行为与记忆洞察生成量化验证规则。 

% Rule Evolution Engine empowers the agent to drive rule iteration through multi-source information fusion, requiring comprehensive utilization of historical rule efficacy (e.g., NDCG fluctuations), real-time group behavioral characteristics (e.g., opinion convergence speed), and insights from the memory module to generate quantitatively validated new rule proposals. The authorization prompt specifies: 

%"You are authorized as a rule repository evolution engine to drive iterative upgrades of the rule system. Based on recent model performance metrics, rule execution efficacy reports, and characteristic features of representative group cases, execute a three-phase optimization: first detect obsolete rule entries; next propose quantifiable new rule candidates requiring computable weight statistical conditions; finally complete version iteration through dual verification standards while synchronously generating change logs documenting rule additions, deletions, and modifications."

% \begin{verbatim}
% 这里是你想要展示的prompt文本
% \end{verbatim}

%规则进化引擎驱动多源信息融合的规则迭代：综合历史规则效能（如NDCG波动）、实时群体行为特征（如观点收敛速度）及记忆模块洞察，生成定量验证的新规则提案。 

\textbf{Updatable Memory Module}.
% DALI框架通过三层协同记忆架构实现认知经验持续优化：版本化规则库采用语义化版本存储可计算规则，更新时生成结构化快照并保留历史版本确保实验复现；案例特征库通过双通道向量化存储归档群体决策模式——领导型群组保留权重集中特征，协作型群组存储分散特性；性能演化日志实时解析训练数据生成轮次粒度指标图谱，以只追加模式记录规则版本效能证据。验证失败的规则自动触发回滚日志保留错误上下文，形成闭环认知进化系统。 
% The DALI framework implements a multi-layered memory architecture for continuous cognitive optimization, comprising three synergistic units: The Versioned Rule Repository stores computable rules with version control, generating snapshots during updates; the Case Feature Bank archives group decision patterns via vectorized storage; the Performance Evolution Log constructs metric maps through real-time parsing, with failed validations triggering rollback logs to establish a closed-loop cognitive system. 
% The \textbf{Versioned Rule Repository} stores computable rules with semantic versioning, generating structured snapshots during updates while preserving historical versions for reproducibility. The \textbf{Case Feature Bank} employs dual-channel vectorized storage to archive group decision patterns—leadership groups retain high weight-concentration features while collaborative groups store decentralized characteristics. The \textbf{Performance Evolution Log} constructs epoch-granular metric maps through real-time parsing, correlating data with rule versions via append-only documentation. Failed validations automatically trigger rollback logs preserving error contexts, establishing a closed-loop cognitive evolution system. 

The DALI framework implements a three-layer, synergistic, updatable memory architecture to enable continuous optimization and evolution of cognitive experience. This module comprises three core units, forming a closed-loop learning system that spans experience storage, pattern archiving, and efficacy tracking:

\textbf{Versioned Rule Repository} Stores all computable rules using semantic versioning. Each update generates a structured snapshot containing metadata and change context, while permanently preserving historical versions to guarantee full experimental reproducibility.
\textbf{Case Feature Bank} Employs a dual-channel vectorized storage strategy to archive representative group decision patterns. Features of leadership-oriented groups are stored with high weight-concentration as the key vector, whereas features of collaborative-oriented groups are stored with a focus on their opinion weight dispersion characteristics, thereby building a searchable repository of decision pattern exemplars.
\textbf{Performance Evolution Log} Constructs epoch-granular, multi-dimensional metric maps by parsing the training data stream in real-time. Operating in an "append-only" mode, this log precisely correlates each fluctuation in performance metrics with the specific rule version that triggered it, forming a continuous chain of efficacy evidence. When a rule validation fails, the system automatically triggers a rollback mechanism and generates a rollback log that preserves the complete error context.

These three units work in concert: the Performance Evolution Log provides empirical evidence for the efficacy of rule versions; the Case Feature Bank offers the pattern basis for understanding which rules are effective under which group modes; and the Versioned Rule Repository undergoes safe iteration driven by the information above. This design ultimately establishes a data-driven, closed-loop cognitive evolution system capable of continuous learning from both historical experience and real-time feedback.

\textbf{Planning Module}. 
%The Planning Module builds a Performance-Rule Feedback Loop (PRFL) for self-iteration: real-time training log monitoring extracts multi-dimensional metrics linked to current rule versions; triggers rule efficacy evaluators quantifying performance drift; activates rule generation-validation pipelines that synthesize "IF-THEN" candidate rules using historical cases and real-time data to complete version iteration. This closed-loop mechanism overcomes static limitations of traditional models, establishing a self-consistent system from experiential accumulation to rule evolution. 

The Planning Module constructs a Performance-Rule Feedback Loop (PRFL) to drive the system's self-iteration and continuous optimization. This module operates by monitoring training logs in real-time to extract multi-dimensional performance metrics linked to the currently active rule version. Upon detecting significant performance drift, it automatically triggers a rule efficacy evaluator for quantitative analysis. This, in turn, activates a rule generation and validation pipeline, which synthesizes new candidate rules with an "IF-THEN" logical structure by integrating historical case features and real-time behavioral data, ultimately completing a secure version iteration of the rule repository. This closed-loop mechanism overcomes the static and lagging limitations inherent in traditional rule systems, establishing a self-consistent cognitive system that enables autonomous evolution from the continuous accumulation of experiential data to the rule system itself. 
% 规划模块构建性能-规则反馈闭环（PRFL）驱动自迭代：实时监控训练日志提取多维指标，关联当前规则版本；触发规则效能评估器量化性能漂移（如关键指标连续下降超阈值）；激活规则生成-验证管道，基于历史案例与实时数据合成"IF-THEN"结构候选规则完成版本迭代。该闭环机制突破传统静态认知局限，建立从经验积累到规则进化的自洽系统。 

\textbf{Action Module}. As DALI's core execution engine, the Action Module achieves precise group dynamics identification and rule optimization via multimodal decisions: first parses high-dimensional weight tensors from user preference encoders, fusing memory module rule bases with LLMs' semantic comprehension to construct a composite analytical framework—matching features through rule engines while invoking LLMs for behavioral parsing to generate rule-based decisions. During rule optimization, it triggers in-memory rule generators based on temporal training data and performance logs, creates candidate rules validated through end-to-end pipelines, then commits verified rules to historical repositories via rule fingerprinting, establishing a closed-loop system from data perception to cognitive iteration.

 \begin{figure*}
  \centering
    \includegraphics[width=1\linewidth,height= 9 cm]
    {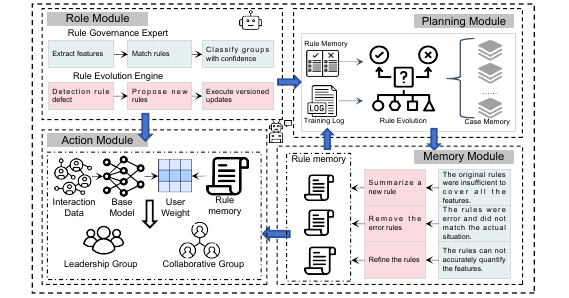}
   \caption{
\textbf{Agent of DALI}. The Role Module establishes the agent's dual roles: Rule Governance Expert (real-time group classification) and Rule Evolution Engine (driving rule iteration). The Planning Module constructs a PRFL loop: detecting obsolete rules, generating new rules with statistical conditions, and executing version iteration for verified rules. The Memory Module stores quantifiable rules. The Action Module loads the rule repository to perform symbolic reasoning and outputs group classification types.}
%  配置模块确立智能体双重角色：规则治理专家（实时群体分类）与规则进化引擎（驱动规则迭代）。规划模块构建PRFL闭环：首先检测失效规则，然后生成带统计条件的新规则，最后经过验证通过的规则进行版本迭代。记忆模块存储量化规则。行为模块加载规则库执行符号推理，输出群体类型。
   \label{fig:structure}
\end{figure*}

\subsection{DALI Training Workflow}

The DALI framework implements an autonomous cognitive closed-loop training process through the synergistic operation of its modular components, which centers on constructing a complete cycle from perception and decision-making to action and learning. As illustrated in Figure 1 (assuming a corresponding diagram), its workflow can be systematically deconstructed into three phases: initialization, collaborative execution, and iterative evolution.

\textbf{Initialization and Role Assignment}: Upon system startup, dual cognitive roles are established. The \textbf{Rule Governance Expert} loads the initial rule repository, preparing for real-time leadership pattern identification. Simultaneously, the \textbf{Rule Evolution Engine} is initialized, defining subsequent rule iteration pathways and decision logic.

\textbf{Collaborative Execution and Computation}: During the execution phase, the \textbf{Operation System} performs symbolic-neural fusion computation on the input data, parsing weight tensors from group interactions to generate preliminary group representations. Concurrently, the \textbf{Rule Governance Expert} applies the rule repository to conduct real-time inference and group classification (leadership/collaborative) based on these representations. The \textbf{Memory Module} provides critical support throughout this process: the \textbf{Versioned Rule Repository} supplies the currently active discrimination logic, the \textbf{Case Feature Bank} offers historical patterns for analogical reasoning, and the \textbf{Performance Evolution Log} records the context of the current inference.

\textbf{Iterative Evolution and Closed-Loop Feedback}: This phase is driven by the Planning Module. Its built-in Performance Rule Feedback Loop (PRFL) continuously monitors the performance metrics of the system output. Upon detecting performance drift (e.g., consecutive deterioration of key metrics), the PRFL triggers an alert and coordinates the \textbf{Rule Evolution Engine} to initiate a rule generation and validation pipeline. Newly generated candidate rules are validated in a sandboxed environment. Successfully validated rules undergo versioned refinement before being submitted for deployment to the rule repository within the \textbf{Memory Module}. If validation fails, a rollback mechanism is triggered to restore a stable version. The updated rules then take effect in the next execution cycle. This constitutes a complete autonomous evolution closed-loop, whose process can be summarized as \textbf{Execution, Monitoring, Evaluation, Evolution, and Re-execution}, forming an iterative cycle. 

% , thereby forming a complete autonomous evolution loop: "Execution -> Monitoring -> Evaluation -> Evolution -> Re-execution."

\subsection{Neuro-Symbolic Hybrid Aggregation Mechanism}

Building upon the LLM-driven agent for leadership identification, this study proposes the Neuro-Symbolic Hybrid Aggregation (NSHA) mechanism within the DALI framework. This mechanism achieves precise recognition of group leadership patterns through \textbf{dual-channel collaborative decision-making}, \textbf{triple-threshold discrimination logic}, and \textbf{attention optimization strategies}. Its core innovation lies in the effective integration of the interpretability and robust reasoning capabilities of symbolic approaches with the representation learning capacity of neural networks, thereby transcending the limitations of traditional single-path methods. 
To ensure accurate pattern identification while mitigating imperfections in the rule base during the early training stages, we construct a synergistic dual-channel decision framework that integrates symbolic reasoning and neural computation. The \textbf{Symbolic Reasoning Channel} leverages the aforementioned LLM-based rule agent to perform dynamic rule-driven pattern recognition. This channel accepts member weight tensors to extract key statistical features, loads the current rule repository for logical inference, and ultimately outputs classification labels accompanied by interpretable rule-matching paths. This process is formalized as: 
\begin{equation}
\begin{gathered}
      \mathbf{P}_{g}^{\text{symbolic}} = \Gamma\left( \mathrm{R}_{\mathrm{v}} \left( W \right) \right) ,
  \end{gathered}
\end{equation}
where $\mathbf{P}_{g}^{\text{symbolic}}$ is the probability matrix, and $\mathrm{R_{V}}$ denotes the dynamic rule agent, and $\Gamma$ is a classification mapping function converting agent outputs into machine-actionable labels (leadership/collaborative group).

% The neural computation channel employs deep learning to capture implicit features through a three-layer gated MLP architecture: accepting identical weight tensors for feature fusion and generating probabilistic classification results:

The \textbf{Neural Computation Channel} employs a three-layer gated Multi-Layer Perceptron (MLP) architecture to capture implicit, complex patterns in the data through deep learning. This channel processes the same input weight tensor W, performs non-linear transformation and feature fusion, and generates a probabilistic classification output: 

 % 第一层：输入层到隐藏层
 % 第二层：隐藏层1到隐藏层2
 % 第三层：隐藏层2到输出层
 % 输出层：Softmax激活
\begin{equation}
\begin{aligned}
    \mathbf{h}^{(1)} = \mathrm{ReLU}\left( \mathbf{W_{1}^{T}} \mathbf{W_{1}} + \mathbf{b_{1}} \right) \\
    \mathbf{h^{(2)}} = \tanh\left( \mathbf{W_{2}^{T}} \mathbf{h^{(1)}} + \mathbf{b_{2}} \right)\\ 
    \mathbf{z} = \mathbf{W_{3}^{T}} \mathbf{h^{(2)}} + \mathbf{b_{3}}\\
    \mathbf{P}_{g}^{\text{neural}} = \mathrm{softmax}\left( \mathbf{z} \right)
\end{aligned}
\end{equation}
where $\mathbf{P}_{g}^{\text{neural}}$ is the probability matrix; $\mathbf{W_{1}}$, $\mathbf{W_{2}}$, $\mathbf{W_{3}}$ and $\mathbf{b_{1}}$, $\mathbf{b_{2}}$, $\mathbf{b_{3}}$ are learnable weight matrices and bias vectors respectively.

This dual-channel design primarily relies on the robustness of the neural channel to guarantee baseline performance during initial training, while the rule base in the symbolic channel continuously evolves, driven by the performance-rule feedback loop. Ultimately, the system integrates the outputs from both channels via an adaptive fusion strategy based on attention weights, yielding a final decision that combines high accuracy with high interpretability.

% where ${P}_{g}^{\text{neural}}$and $P_{g}^{\text{symbolic}}$ is the probability matrix; $W_{1}$, $W_{2}$, $W_{3}$ and $b_{1}$, $b_{2}$, $b_{3}$ are learnable weight matrices and bias vectors respectively.

\subsection{Triple-Threshold Decision Logic
}
% After obtaining leadership group discrimination results through dual channels, the model returns these outcomes to the attention module and implements differentiated attention aggregation mechanisms based on group type. When dual-channel discrimination results are consistent, the model naturally determines that the outcomes accurately reflect whether a group is a leadership group,

After obtaining leadership group discrimination results through the neural and symbolic dual channels, the model returns these outcomes to the attention module and implements differentiated attention aggregation mechanisms based on the identified group type.

\textbf{Consistent Dual-Channel Results}: When the discrimination results from the neural and symbolic channels are consistent, the model deems the result reliable for reflecting the actual group type (leadership or collaborative) and adopts it directly:
\begin{equation}
\begin{gathered}
P_{g} = P_{g}^{\mathrm{neural}} = P_{g}^{\mathrm{symbolic}}.
  \end{gathered}
\end{equation}
% $$ P_{g} = P_{g}^{\mathrm{neural}} = P_{g}^{\mathrm{symbolic}} $$
% If a group is identified as a leadership group, the preference of the member with the highest weight (i.e. the leader) replaces the group preference,

\textbf{Processing for Leadership Groups}: If a group is identified as a leadership group, the preference of the member with the highest weight (i.e., the leader) solely represents the collective preference of the group: 

% If a group is identified as a leadership group, the preference of the member with the highest weight (i.e. the leader) replaces the group preference,

\begin{equation}
\begin{gathered}
p_{g}^{\mathrm{leader}} = p_{u_{0}} ,
  \end{gathered}
\end{equation}
\begin{equation}
\begin{gathered}
u_{0} = \arg\max_{1 \leq u \leq n} w_{u} ,
  \end{gathered}
\end{equation}
where $w_{u}$ denotes member weights and n is the group size. 

% For leadership groups, DALI preserves the base group recommendation model's attention aggregation mechanism, formalized as:
\textbf{Processing for Collaborative Groups}: For collaborative groups, DALI preserves the attention aggregation mechanism of the base group recommendation model, where the group preference is the weighted sum of individual member preferences based on attention weights: 
\begin{equation}
\begin{gathered}
p_{g}^{\mathrm{collab}} = \sum_{u=1}^{n} \alpha_{u} \cdot p_{u}.
  \end{gathered}
\end{equation}
% $$ P_{g}^{\mathrm{collab}} = \sum_{u=1}^{n} \alpha_{u} \cdot P_{u}^{\psi} $$
% If dual-channel results conflict, a fusion strategy linearly combines both outputs,

\textbf{Fusion Strategy for Conflicting Results}: When the discrimination results from the two channels conflict, the model employs a linear fusion strategy that combines the outputs from both sources:

\begin{equation}
\begin{gathered}
P_{g} = \gamma P_{g}^{\mathrm{symbolic}} + (1 - \gamma) P_{g}^{\mathrm{neural}}.
  \end{gathered}
\end{equation}
% $$ P_{g} = \gamma P_{g}^{\mathrm{symbolic}} + (1 - \gamma) P_{g}^{\mathrm{neural}} $$
% The mixing coefficient $\gamma$ progressively decays with training epochs:

The mixing coefficient $\gamma$ dynamically decays as training progresses (with increasing epoch number). This design aims to place greater trust in the inductive bias of symbolic reasoning during initial training phases, gradually shifting reliance towards the neural model, which learns more comprehensively from data in later stages: 

\begin{equation}
\begin{gathered}
\gamma = \max\left(0,\ 1 - \frac{t}{t_{\max}}\right),
  \end{gathered}
\end{equation}
% $$ \gamma = \max\left(0,\ 1 - \frac{t}{t_{\max}}\right) $$
where $t$ is the current epoch and $t_{max}$  is the maximum training epoch.

\subsection{Model Training Protocol}

This framework adopts a multi-phase collaborative training strategy, achieving efficient model convergence through pretraining initialization and joint optimization.

\textbf{Phase 1: Pretraining}.
% (User Preference Encoder Initialization)
% This phase freezes group aggregation modules to focus on user preference modeling. Aligning with base group recommendation models, DALI refrains from agent involvement during pretraining, utilizing the user training methodology of foundational group recommenders.

In this phase, all group aggregation modules are frozen to focus exclusively on user preference modeling. Aligning with the methodology of base group recommenders, the DALI framework does not involve the agent module during pretraining, instead fully adopting the user training procedure of the foundational recommender to learn high-quality user representations.

 \textbf{Phase 2: Joint Optimization}.
%  (Dual-Channel Collaborative Training)
In the dual-channel collaborative training framework, the system integrates objectives from both base group recommendation and leadership group identification. We mainly consider two different loss functions, $L_{\text{group}} $ and $L_{\text{weight}} $, where $L_{\text{group}} $ optimizes overall recommendation performance inherited from the base model. Specifically, $L_{\text{group}} $ comprises the multi-class cross-entropy loss for implicit feedback at both the group and user levels in the group recommendation system, as well as a mutual information loss between members and groups designed via a binary cross-entropy loss. 

% $$ L_{\text{joint}} = L_{\text{group}} + L_{\text{weight}} $$
% where $L_{group}$ optimizes overall recommendation performance inherited from the base model.

 $L_{\text{weight}} $ verifies that identified leadership groups contain dominant members. This loss ensures significant attention weight concentration for leaders in leadership groups while maintaining balanced distributions in collaborative groups, achieved through dynamic benchmark comparison and type-aware optimization.

The mechanism operates as follows: First, identify all leadership groups in the current batch  and compute each group's dominance score, 
\begin{equation}
\begin{gathered}
\eta = \frac{w_{\max}^{(g)} - \mu_{\text{rest}}^{(g)}}{\mu_{\text{rest}}^{(g)}}.
  \end{gathered}
\end{equation}
% $$ \eta_{g} = \frac{w_{\max}^{(g)} - \mu_{\text{rest}}^{(g)}}{\mu_{\text{rest}}^{(g)}} $$
Then sample K collaborative groups randomly from the batch to establish a collaborative benchmark, 
\begin{equation}
\begin{gathered}
\xi = \frac{1}{K} \sum_{k=1}^{K} \eta^{(k)}.
  \end{gathered}
\end{equation}
% $$ \xi = \frac{1}{K} \sum_{k=1}^{K} \eta^{(k)} $$

Compute the relative dominance for leadership groups, 
\begin{equation}
\begin{gathered}
\Delta = \eta_{g} - \eta_{\text{base}} .
  \end{gathered}
\end{equation}

% $$ \Delta \eta_{g} = \eta_{g} - \eta_{\text{base}} $$
Impose weight loss on leadership groups with insufficient dominance, 
\begin{equation}
\begin{gathered}
 L_{\text{weight}} = \frac{1}{|B_{L}|} \sum_{g \in B_{L}} \max \left(0, \delta - \Delta \right).
  \end{gathered}
\end{equation}
% $$ L_{\text{weight}} = \frac{1}{|B_{L}|} \sum_{g \in B_{L}} \max \left(0, \delta - \Delta \right) $$

Thereafter, we combined the two types of losses together, 
\begin{equation}
\begin{gathered}
%L_{\text{joint}} = L_{\text{group}} + L_{\text{weight}} .
L = L_{\text{group}} + L_{\text{weight}} .
  \end{gathered}
\end{equation}
This approach establishes dynamic thresholds via real-time collaborative group sampling, avoiding distribution bias from fixed thresholds. It correctly models weight disparities in leadership groups while preserving equilibrium in collaborative groups, reinforcing leader dominance and maintaining member equality.

\section{Experiment}\label{}

\begin{table}[t]\normalsize

  \centering
\caption{Ablation Study on DALI.}

    \small
    \fontsize{7.5}{8}\selectfont
    \begin{tabular}{lccccc}\toprule
        Dataset&   Users& Groups& Items&User-Item&Group-Item\\\midrule
 
CAMRa2011& 602& 290& 7,710& 116,344&145,068\\

Mafengwo& 5,275& 995& 1,513& 39,761&3,595\\\bottomrule
    \end{tabular}
    
\label{tab-sota-rgbt}
  
\end{table}

In the experimental section, we will address the following questions. 
RQ1: Can the current framework seamlessly integrate into the basic model and enhance its performance?
RQ2: Can each component of the model independently enhance the model's performance?
RQ3: Whether the different batch sizes will affect the efficiency of rule generation in the model? 
RQ4: Do the rules generated by the model accurately reflect the leadership rules of the group?

\begin{table*}[t] % 使用table*环境实现跨双栏
\centering
\caption{Overall Performance comparison on Mafengwo dataset.}
\label{tab-sota-rgbt}
\small % 统一设置字号
\setlength{\tabcolsep}{6pt} % 减小列间距
% \begin{tabular}{@{}lccccc@{}} % @{}消除左右边距
% \toprule
% Method & HR@5 & HR@10 & NDCG@5 & NDCG@10 & \makecell{Avg.Improv.} \\
% 修改列定义：所有列左对齐
\begin{tabular}{@{}l *{5}{l}@{}} % 使用 *{5}{l} 表示5个左对齐列
\toprule
Method & 
\multicolumn{1}{l}{HR@5} & 
\multicolumn{1}{l}{HR@10} & 
\multicolumn{1}{l}{NDCG@5} & 
\multicolumn{1}{l}{NDCG@10} & 
% \multicolumn{1}{l}{\makecell{Avg.Improv.}} \\
\multicolumn{1}{l}{{Avg.Improv.}} \\
\midrule

% \rowcolor{blue!10} % 添加行色突出
GroupIM       & 0.579      & 0.692      & 0.440      & 0.475      & - \\
GroupIM+LARGE & 0.617 ↑6.6\%  & 0.717 ↑3.6\% & 0.472 ↑7.2\% & 0.504 ↑6.1\% & +5.9\%\\ 
GroupIM+DALI  & \textbf{0.652} ↑12.4\%& \textbf{0.755} ↑9.1\%& \textbf{0.511} ↑16.1\%& \textbf{0.545} ↑14.5\%& \textbf{+13.0\%}\\

\cline{1-1}

% \rowcolor{blue!10}
AGREE         & 0.701      & 0.784      & 0.582      & 0.609      & - \\
AGREE+LARGE & 0.764 ↑9.0\% & 0.831 ↑6.0\% & 0.606 ↑4.1\% & 0.615 ↑1.0\% & +5.0\%\\
AGREE+DALI    & \textbf{0.830} ↑18.3\% & \textbf{0.876} ↑11.8\% & \textbf{0.735} ↑26.3\% & \textbf{0.750} ↑23.2\% & \textbf{+19.9\%} \\

\cline{1-1}

% \rowcolor{blue!10}
HCR           & 0.776      & 0.833      & 0.698      & 0.716      & - \\
HCR+LARGE  & \textbf{0.792} ↑2.0\% & \textbf{0.841} ↑1.0\% & 0.701 ↑0.4\% & 0.714 ↓0.3\% &  \textbf{+0.8\%}\\
HCR+DALI      & 0.785 ↑1.2\% & 0.830 ↓0.3\% & \textbf{0.705} ↑1.0\% & \textbf{0.720} ↑0.4\% & +0.6\%\\

\cline{1-1}

% \rowcolor{blue!10}
CubeRec                 & 0.855            & \textbf{0.901}            & 0.756            & 0.772             & - \\
CubeRec+LARGE  & 0.857 ↑0.2\% & 0.892 ↓1.0\% & 0.754 ↓0.3\% & 0.771 ↓0.1\% & -0.3\% \\
CubeRec+DALI     & \textbf{0.858} ↑0.4\% & 0.893 ↓0.9\% & \textbf{0.765} ↑1.1\% & \textbf{0.776} ↑0.6\% & \textbf{+0.3\%} \\

\cline{1-1}

 %\rowcolor{blue!10}
DisRec        & 0.716 & 0.797 & 0.589 & 0.617 & - \\
DisRec+LARGE & 0.723 ↑1.0\% & 0.811 ↑1.8\% & 0.604 ↑2.6\% & 0.636 ↑3.1\% & 2.1\% \\
DisRec+DALI   & \textbf{0.729} ↑1.9\% & \textbf{0.820} ↑2.9\% & \textbf{0.626} ↑6.1\% & \textbf{0.655} ↑6.3\% & \textbf{+4.3\%} \\

\bottomrule
\end{tabular}
\par \smallskip
% \footnotesize\textit{Note: ↑/↓ indicate performance change, improvements in bold}
\end{table*}
% 表格再调整一下，数字对齐！

\subsection{Experimental setup}

\textbf{Datasets.}
% We conducted experiments on three real-world datasets. The Mafengwo dataset is a tourism website dataset that contains individual user and group travel records. We randomly divided the groups into training set (60\%), validation set (20\%), and test set (20\%). 

We conduct experiments on two real-world datasets: CAMRa2011, a movie rating dataset with individual and home users, and Mafengwo, a travel site with individual and group travel records. Dataset details are in Table 1. We randomly divided the groups into training set (60\%), validation set (20\%), and test set (20\%).

\textbf{Experimental Environment.}
All experiments were conducted on a Linux system (Ubuntu 22.04 LTS) equipped with an Intel® Core™ i9-14900K CPU, 64GB RAM, and an NVIDIA GeForce RTX 4090 GPU (24GB VRAM), utilizing Python 3.10.16 with PyTorch 1.13.1 and NumPy 1.26.4. 

\textbf{Baseline models.} Our framework can be applied to most existing group recommendation models to enhance their performance, demonstrating excellent universality. We combined our framework with some representative group recommendation models.

AGREE \cite{cao2018attentive} incorporates attention mechanisms for group modeling; GroupIM \cite{sankar2020groupim} captures member contributions via mutual information maximization; HCR \cite{jia2021hypergraph} constructs hypergraph-based member relationships; CubeRec \cite{chen2022thinking} models groups using hypercubes with distance metrics; DisRec \cite{ye2025disentangled} enhances group modeling through social relationship integration; LARGE \cite{gan2025large} uncovers leadership structures with influential members dominating decisions.

 \subsection{Overall performance comparison (RQ1)}
 
% In this section, we will demonstrate the outstanding performance of the proposed DALI algorithm on the mafengwo dataset with Table 1. Method + DALI represents the integrated version of the DALI algorithm combined with the base GR model. Based on these results, several key observations can be drawn.

\begin{table*}[t] % 使用table*环境实现跨双栏
\centering
\caption{Overall Performance comparison on CAMra2011 dataset.}
\label{tab-sota-rgbt}
\small % 统一设置字号
\setlength{\tabcolsep}{6pt} % 减小列间距
% \begin{tabular}{@{}lccccc@{}} % @{}消除左右边距
% \toprule
% Method & HR@5 & HR@10 & NDCG@5 & NDCG@10 & \makecell{Avg.Improv.} \\
% 修改列定义：所有列左对齐
\begin{tabular}{@{}l *{5}{l}@{}} % 使用 *{5}{l} 表示5个左对齐列
\toprule
Method & 
\multicolumn{1}{l}{HR@5} & 
\multicolumn{1}{l}{HR@10} & 
\multicolumn{1}{l}{NDCG@5} & 
\multicolumn{1}{l}{NDCG@10} & 
% \multicolumn{1}{l}{\makecell{Avg.Improv.}} \\
\multicolumn{1}{l}{{Avg.Improv.}} \\
\midrule

% \rowcolor{blue!10} % 添加行色突出
GroupIM       & 0.561      & 0.752      & 0.377      & 0.439      & - \\
GroupIM+LARGE & 0.557 ↓0.7\%   & 0.752 ↑0.0\%  & 0.370 ↓1.9\%  & 0.434 ↓1.1\%   & -0.9\%\\
GroupIM+DALI  & \textbf{0.581} ↑3.6\%& \textbf{0.761} ↑1.2\%& \textbf{0.386} ↑2.4\%& \textbf{0.445} ↑1.4\%& \textbf{+2.2\%}\\

\cline{1-1}

% \rowcolor{blue!10}
AGREE         & 0.859      & 0.890      & 0.844      & 0.854      & - \\
AGREE+LARGE & 0.856 ↓0.3\% & 0.888 ↓0.2\% & 0.839 ↓0.6\% & 0.848 ↓0.7\% & -0.5\%\\
AGREE+DALI    & \textbf{0.926} ↑7.8\% & \textbf{0.936} ↑5.2\% & \textbf{0.904} ↑7.1\% & \textbf{0.907} ↑6.2\% & \textbf{+19.9\%} \\

\cline{1-1}

% \rowcolor{blue!10}
HCR           & 0.592      & 0.775      & 0.472      & 0.531      & - \\
HCR+LARGE  & 0.602 ↑1.7\% & 0.714 ↓7.9\% & 0.554 ↑17.4\% & 0.604 ↑13.7\% &  +6.2\%\\
HCR+DALI      & \textbf{0.613} ↑3.5\% & \textbf{0.763} ↓1.5\% & \textbf{0.565} ↑19.7\% & \textbf{0.613} ↑15.4\% & \textbf{+9.3\%}\\

\cline{1-1}

% \rowcolor{blue!10}
CubeRec                 & 0.585            & 0.784            & 0.380            & 0.444             & - \\
CubeRec+LARGE  & 0.590 ↑0.9\% & \textbf{0.786} ↑0.3\% & 0.373 ↓1.9\% & 0.440 ↓0.9\% & -0.4\% \\
CubeRec+DALI     & \textbf{0.591} ↑1.0\% & 0.782 ↓0.3\% & \textbf{0.382} ↑0.5\% & \textbf{0.451} ↑1.6\% & \textbf{+0.3\%} \\

\cline{1-1}

 %\rowcolor{blue!10}
DisRec        & 0.875 & 0.890 & 0.862 & 0.867 & - \\
DisRec+LARGE & 0.876 ↑0.1\% & 0.886 ↓0.4\% & 0.872 ↑1.2\% & 0.870 ↑0.3\% & +0.3\% \\
DisRec+DALI   & \textbf{0.891} ↑1.8\% & \textbf{0.912} ↑2.5\% & \textbf{0.884} ↑2.6\% & \textbf{0.891} ↑2.8\% & \textbf{+2.4\%} \\

\bottomrule
\end{tabular}
\par \smallskip
% \footnotesize\textit{Note: ↑/↓ indicate performance change, improvements in bold}
\end{table*}

This section systematically evaluates the performance of the proposed DALI framework through experiments on two datasets: Mafengwo and CAMra2011. Tables 2 and 3 present the comparative results of various base group recommendation (GR) models, their integrated versions with the alternative leadership-aware framework LARGE, and their integrated versions with DALI on key metrics. Here, "Method+DALI" denotes the version combining the DALI algorithm with the base GR model. Based on the experimental results, the following key observations can be drawn:

\textbf{DALI Significantly Enhances Diverse Base Models}: Experimental analysis demonstrates that the DALI framework significantly enhances the performance of diverse GR models. For foundational models (e.g., GroupIM), DALI delivers substantial gains in key metrics (Avg.Improv. +13.0\%). For attention-based models (e.g., AGREE), DALI achieves breakthrough improvements (Avg.Improv. +19.9\% on Mafengwo and +19.9\% on CAMra2011), validating its efficacy in modeling leadership groups via its neuro-symbolic hybrid mechanism to overcome the limitations of conventional attention. Even when integrated with high-performance state-of-the-art models (e.g., HCR, CubeRec, DisRec), DALI exhibits consistent positive enhancements, yielding significant gains for DisRec (Mafengwo: +4.3\%, CAMra2011: +2.4\%), which confirms its strong compatibility and enhancement capability.
\textbf{Superiority of DALI over the LARGE Framework}: The comparative experiments with the alternative leadership-aware framework LARGE show that DALI outperforms LARGE in the vast majority of cases. On foundational models like GroupIM and AGREE, the improvements brought by DALI (Mafengwo: +13.0\%/+19.9\%) far exceed those of LARGE (+5.9\%/+5.0\%). Crucially, on the CAMra2011 dataset, LARGE even leads to performance degradation when combined with GroupIM and AGREE (Avg.Improv. -0.9\%/-0.5\%), whereas DALI maintains stable improvements (+2.2\%/+19.9\%). This highlights the robustness of DALI's neuro-symbolic fused architecture in accurately identifying dynamic leadership patterns.
\textbf{Robustness with High-Performing Baselines}: When the base model itself already exhibits high performance (e.g., CubeRec, HCR), the room for further improvement diminishes. Nevertheless, DALI consistently achieves non-negative gains and generally outperforms LARGE on key ranking quality metrics (NDCG@5/10). For instance, on Mafengwo, CubeRec+DALI improves NDCG@5 by 1.1\%, while the LARGE version decreases it by 0.3\%. This validates the effectiveness of DALI's dual-channel adaptive architecture, which can not only empower foundational models with transformative leaps but also provide compatible and robust enhancements for advanced models.
\textbf{Cross-Dataset Robustness}: DALI demonstrates a consistent performance improvement trend across two datasets with different characteristics (Mafengwo and CAMra2011), proving the generalization ability of its framework design. Despite the already high absolute performance of some base models on CAMra2011, DALI still manages to deliver further significant gains (e.g., +19.9\% for AGREE+DALI), strongly supporting its potential for application in diverse group decision-making scenarios.

In conclusion, the experimental results robustly validate the technical superiority of the DALI framework from multiple dimensions. Its neuro-symbolic hybrid architecture and dual-channel decision logic enable more accurate identification of dynamic leadership patterns and adaptive adjustment of group preference aggregation strategies, leading to significant and consistent performance enhancements across a wide range of base models.

\begin{table}[t]\normalsize

  \centering
\caption{Ablation Study on DALI.}

    \small
    \fontsize{7.5}{8}\selectfont
    \begin{tabular}{lcccc}\toprule
        Method&   HR@5& HR@10& NDCG@5&NDCG@10\\\midrule
 
Group IM-DALI    & \textbf{0.652 }& \textbf{0.755}& \textbf{0.511}& \textbf{0.545 }\\

Group IM-A    & 0.626& 0.727& 0.493& 0.526\\

Group IM-N    & 0.617& 0.717& 0.471& 0.504\\

Group IM-S    & 0.613& 0.717& 0.476& 0.510\\

GroupIM       & 0.579& 0.692& 0.438& 0.475\\

    \bottomrule
    \end{tabular}
    
\label{tab-sota-rgbt}
  
\end{table}

\subsection{Analysis of Cross-Dataset Performance Variance}
The variance in performance gains achieved by the DALI framework between the Mafengwo and CAMRa2011 datasets can be reasonably explained by analyzing the intrinsic characteristics of the datasets. As shown in Table 1, the two datasets differ fundamentally in group structure, which directly affects the utility boundary of enhancement methods based on leadership identification.

The core reason lies in the \textbf{significant difference in average group size} and the consequent variation in group decision-making dynamics. The CAMRa2011 dataset exhibits an extremely dense "Group-Item" interaction record (145,068), yet a relatively small number of groups (290). This suggests its groups likely represent small, tightly-knit, and stable collectives (e.g., families, close friend circles). Within such small-scale, homogeneous groups, the decision-making process tends to be more collaborative, with smaller preference disparities among members, making it difficult for a single member with absolute weight dominance (a "leader") to emerge or be necessary. Therefore, the potential gain space for DALI's mechanism, which is designed to precisely identify and leverage leadership patterns, is inherently limited. 

Conversely, the group structure of the Mafengwo dataset aligns better with DALI's design assumptions. Its "Group-Item" interactions are relatively sparse (3,595), but the number of groups is larger (995), more closely resembling the scenario of large-scale, ad-hoc interest groups commonly formed for specific activities on online social platforms. Within such groups, due to the larger member count and potentially looser interest overlap, the decision-making process is more prone to being dominated by a few active or authoritative members, meaning leadership patterns are more pronounced. DALI's neuro-symbolic hybrid mechanism—particularly the rule-based reasoning of the symbolic channel and the triple-threshold logic—is explicitly designed to effectively isolate and amplify leadership signals in such complex, dynamic, large-scale groups. Consequently, it achieves more substantial performance improvements on Mafengwo.

This analysis not only explains the cross-dataset performance variance but also further \textbf{validates the rationale and advantageous scenarios of the DALI methodology}. The results indicate that DALI is not a universal enhancer agnostic to group structure, but rather a framework specifically optimized for \textbf{large-scale or heterogeneous groups where significant leadership dynamics exist}. Its performance gain is positively correlated with the clarity of the internal power structure within a group, which provides strong evidence confirming the effectiveness of its core mechanism (leadership pattern identification and differentiated aggregation) and its practical application value. 

\subsection{Ablation Study (RQ2)}

To validate the efficacy of core components in the DALI framework, this study designs a systematic ablation experiment on the Mafengwo dataset, comparing the full framework (GroupIM-DALI) against three critical variants:
GroupIM-A removes the neural discriminator module to isolate its contribution; GroupIM-N eliminates the rule-based agent module to assess rule-learning impact; GroupIM-F employs exclusively frozen rules for leadership group identification, testing dynamic rule evolution necessity.

All experiments adopt identical training configurations (Adam optimizer, lr=0.001, batch size=32, 50 epochs), with primary evaluation focusing on recommendation metrics (NDCG@5/10, Hit@5/10). The experimental results are presented in Table 2.

 Experimental results of Table 2 conclusively demonstrate that the synergistic design of DALI’s neuro-symbolic hybrid architecture is pivotal for performance enhancement. The full framework significantly outperforms all variants across every metric, validating the complementary value of the dynamic rule engine and neural discriminator. Performance degradation in GroupIM-A underscores the irreplaceable capability of neural components in capturing implicit decision patterns. The most substantial decline observed in GroupIM-N confirms that the rule agent critically extracts key patterns from leadership groups, enabling accurate modeling of their characteristics. Notably, the performance drop in GroupIM-F highlights the necessity of the dynamic rule evolution mechanism, as manually designed frozen rules fail to generalize complex representations of group power structures.

\begin{table}[t]\normalsize

  \centering
\caption{Hyperparameter Sensitivity Experiment on DALI.}

    \small
    \fontsize{7.5}{8}\selectfont
    \begin{tabular}{lcccc}\toprule
        Batch\_size&   HR@5& HR@10& NDCG@5&NDCG@10\\\midrule
 
16& 0.626& 0.727& 0.493& 0.526\\

32& \textbf{0.652 }& \textbf{0.755}& \textbf{0.511}& \textbf{0.545 }\\

64& 0.649& 0.744& 0.512& 0.542\\

128& 0.651& 0.745& 0.515& 0.545\\

256& 0.624& 0.731& 0.492& 0.527\\

    \bottomrule
    \end{tabular}
    
\label{tab-sota-rgbt}
  
\end{table}

 \subsection{Hyperparameter Sensitivity Experiment (RQ3)}
 %实验系统探究了批次规模对DALI框架性能的影响机制。在Mafengwo数据集上固定学习率（lr=0.001）与丢弃率（dropout=0.4）等超参数，测试批次规模从16至256的模型表现。实验数据显示，当batch\_size=32时模型达到综合性能峰值：Hit@5为0.65226，NDCG@5达0.51125。小批次（16）引发显著性能衰减，Hit@10降至0.72663，NDCG@10波动加剧；大批次（256）同样导致性能劣化，Hit@5跌至0.62412，NDCG@10衰减至0.52723。这一现象可归因于批次规模的双重作用机制：小批次因梯度估计方差过高导致规则学习失稳，使LLM智能体难以识别稳定的领导模式特征；大批次则因梯度估计偏差过低，削弱模型对不准确规则的修正能力，使其陷入次优收敛状态。理想批次区间（32-128）恰好平衡偏差-方差矛盾，既保障规则特征提取的稳定性，又保留对错误规则的迭代优化空间。基于此，提出工程实践建议：在算力时间充足场景优先采用batch\_size=32以实现最优精度；最好严格避免极端批次规模以防训练失稳。该结论为资源差异化部署提供理论依据，证实DALI框架在32-128批次区间具备鲁棒适应性。  

\begin{figure}
  \centering
  \includegraphics[width=0.8\linewidth,height= 4 cm]
    {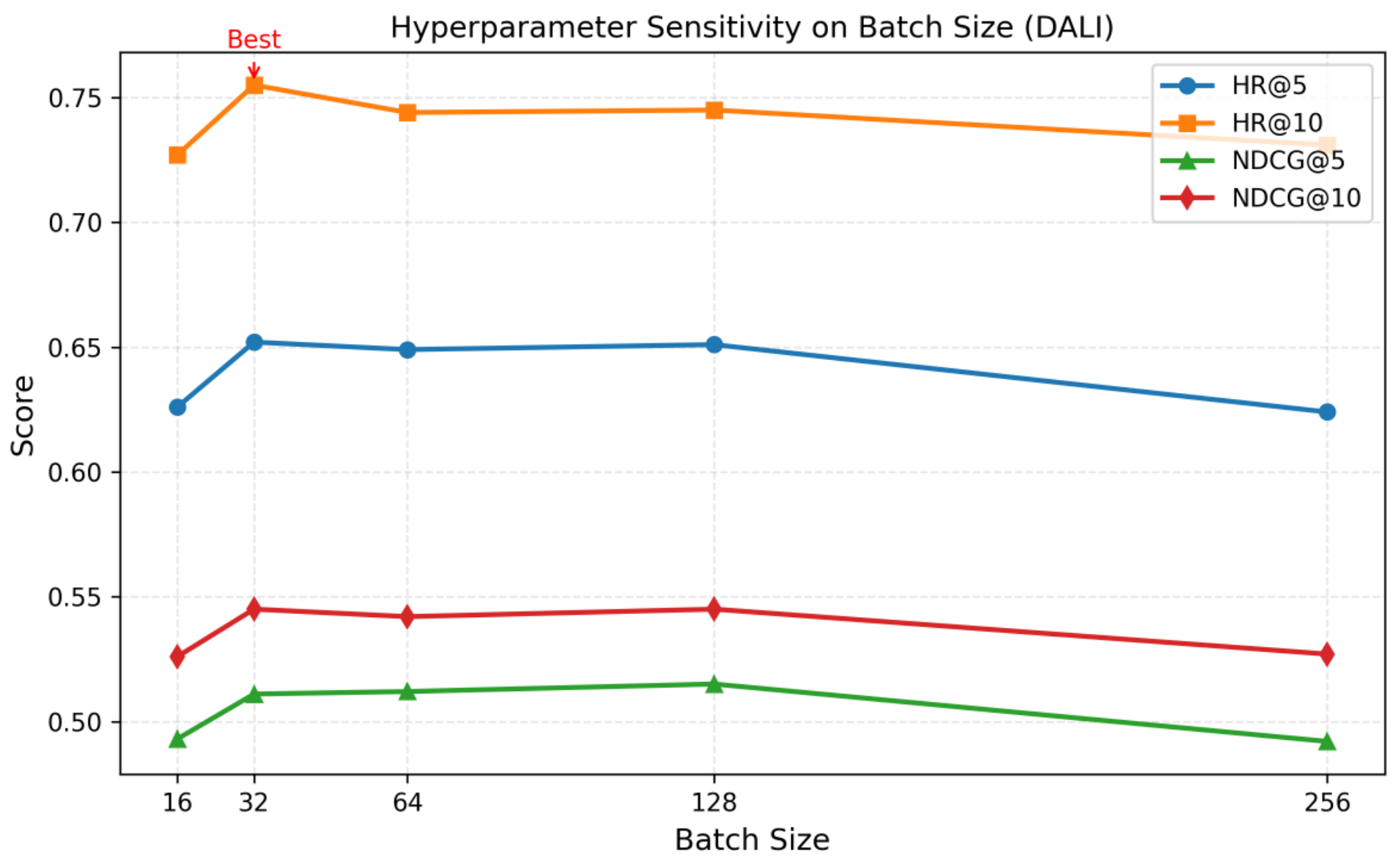}
   \caption{
Hyperparameter Sensitivity on Batch Size.}
   \label{fig:structure}
\label{fig3}
\end{figure}

Within the DALI framework, the model requires leadership group identification within each batch, necessitating an investigation into the impact of batch size variations on framework performance. This experiment systematically examines the influence of batch scale on DALI's operational efficacy. Conducted on the Mafengwo dataset with fixed hyperparameters (learning rate lr=0.001, dropout rate=0.4), the study evaluates batch sizes ranging from 16 to 256. The experimental results are presented in Table 3.
% The DALI framework's leadership group identification requires batch-level processing, necessitating examination of batch size impact on performance. Experiments on the Mafengwo dataset with fixed hyperparameters (lr=0.001, dropout=0.4) tested batch sizes from 16 to 256. The experimental results are presented in Table 3.

% Experimental results demonstrate that the model achieves peak comprehensive performance at batch\_size=32: Hit@5 reaches 0.652 while NDCG@5 attains 0.511. Smaller batches (16) induce significant performance degradation, reducing Hit@10 to 0.727 and intensifying NDCG@10 fluctuations. Conversely, larger batches (256) cause performance deterioration, with Hit@5 declining to 0.624 and NDCG@10 decaying to 0.527.
% Results show peak performance at batch\_size=32 (Hit@5=0.652, NDCG@5=0.511). Smaller batches (16) cause significant degradation (Hit@10=0.727) and unstable NDCG@10 fluctuations. Larger batches (256) reduce performance (Hit@5=0.624, NDCG@10=0.527), demonstrating the critical role of batch size tuning for optimal framework efficacy. 
Results show peak performance at batch\_size=32. Smaller batches (16) cause significant degradation and unstable NDCG@10 fluctuations. Larger batches (256) reduce performance, demonstrating the critical role of batch size tuning for optimal framework efficacy. 

This phenomenon is attributed to the dual mechanism of batch scale effects: Small batches exhibit high gradient estimation variance, destabilizing rule learning and impeding the LLM agent's ability to identify consistent leadership patterns. Large batches suffer from low gradient estimation bias, diminishing the model's capacity to correct inaccurate rules and trapping optimization in suboptimal convergence. The optimal batch range (32-128) balances the bias-variance trade-off, ensuring both stable feature extraction for rule formulation and preserved iterative optimization for erroneous rule correction.

% Based on these findings, engineering implementation guidelines recommend: Prioritizing batch size=32 in computationally resource-sufficient scenarios for maximal accuracy.  Strictly avoiding extreme batch sizes to prevent training instability. 

% This conclusion provides theoretical foundations for resource-heterogeneous deployment strategies, confirming DALI's robust adaptability within the 32-128 batch size range.

\subsection{Case Study (RQ4)}

% To validate DALI-generated rules' accurate reflection of leader group dynamics, we analyze three critical updates. For abrupt decision groups (e.g., children dominating attraction selection), epoch 1-2 detected group NDCG@10 plunge 0.039 triggering epoch 3 rule generation; regarding power-transition groups (e.g., corporate leadership handover), continuous decline 0.006 during epoch 6-7 prompted rule evolution; for deadlock-breaking groups (e.g., family restaurant conflicts), epoch 2 data showed group NDCG@10 drop 0.039 with 0.110 loss increase, leading to rule enhancement. The generated rule system is: 

To validate DALI-generated rules' reflection of leader group dynamics, we analyze three critical updates: For abrupt-decision groups, epoch 1-2 group NDCG@10 plunge 0.039 triggered rule generation; for power-transition groups, epoch 6-7 continuous decline 0.006 prompted evolution; for deadlock-breaking groups, epoch 2 data showed group NDCG@10 drop 0.039 with 0.110 loss increase and drove enhancement. The rule system is: 

\begin{tcolorbox}[colback=gray!10, colframe=gray!50, rounded corners, boxrule=0.5pt, left=3pt,  right=3pt,  top=2pt, bottom=2pt, before skip=5pt, after skip=5pt, fontupper=\small ]

\textbf{Abrupt Decision Rule}:

If $group\_ndcg@10$ single-epoch fluctuation exceeds 0.01, trigger rollback and record leader weight characteristics.

\textbf{Power Transition Rule}:

When $group\_ndcg@10$ decreases by more than 0.005 for three consecutive epochs, automatically lower the $user\_ndcg@10$ rollback threshold to 0.001.

\textbf{Deadlock Resolution Rule}:

If $group\_ndcg@10$ decreases by over 0.01 with \textit{u}\textit{ser}\_\textit{n}\textit{d}\textit{c}\textit{g}@10 change below 0.001, and either loss decreases by exceeds  0.4 or single-epoch fluctuation exceeds  0.01, trigger intervention.

\end{tcolorbox}

Empirical results confirm: The abrupt decision rule captured two similar events in epoch 3-4, validating effective modeling of explosive power concentration; the power transition rule narrowed subsequent epoch 7-8 fluctuations to 0.003, proving precise response to gradual shifts; the deadlock resolution rule prevented false triggers in epoch 6 data by distinguishing normal declines from pre-deadlock phases. Tripartite case studies demonstrate LLM-driven agents precisely extract correlations between evaluation metrics and leadership rules—associations difficult for manual rule design to capture.

\section{Related Work}\label{}

\subsection{Learning-based Group Recommendation Systems}

% Early approaches universally assumed all members influence decisions equally \cite{deng2021knowledge}, manifesting in two paradigms: Static rule aggregation \cite{malecek2021fairness, sato2022enumerating, stratigi2023squirrel} ignores internal power structures, introducing systematic bias; Dynamic learning models \cite{huang2020efficient, wu2023consrec} capture member interactions but retain the universal contribution assumption. 

Early studies relied on predefined, static heuristic rules for aggregating group preferences, such as the average strategy (Average)  \cite{deng2021knowledge} and the least misery strategy (Least Misery) \cite{alvarado2022systematic}. Static rule aggregation \cite{malecek2021fairness, sato2022enumerating, stratigi2023squirrel} ignores internal power structures, introducing systematic bias; Dynamic learning models \cite{huang2020efficient, wu2023consrec} capture member interactions but retain the universal contribution assumption. 
Although these methods were simple and intuitive, they were unable to adapt to dynamic and variable group decision-making scenarios. With the development of deep learning, research shifted to data-driven neural network methods, achieving a paradigm shift from "static aggregation" to "dynamic aggregation". The introduction of the attention mechanism (such as the AGREE model) \cite{cao2018attentive} marked a significant advancement, enabling dynamic learning of the relative importance of each member in the decision-making process. MoSAN \cite{vinh2019interact} and other similar methods further adopted more complex sub-attention networks to capture subtle interactions.
SSSADRGR\cite{krishnamoorthi2026self} discovers rich social connections and contextual information within social networks to achieve personalized and group-aware recommendations. It learns group preferences through self-supervised learning from internal social interactions and group dynamics.
These methods typically represent the group as a single point embedding vector, which has inherent flaws. It forcibly compresses diverse and complex group preferences into a fixed dimension, inevitably leading to information loss and making it difficult to accurately represent the differential distribution of preferences within the group, let alone simulate the decision-making process in reality where the group reaches an "acceptable range" through negotiation. \cite{chen2022thinking}

To overcome the limitations of point embeddings, the research has been deepened in two directions. On the one hand, the development of graph neural networks (GNNs) has made the modeling of complex relationships clearer. \cite{gao2025graph, wang2026graph } Some recent approaches focus on using graph neural networks to model groups, such as in works like SGGCF \cite{li2023self} and MMAN \cite{yin2023beyond}, which construct heterogeneous graphs of users, groups, and items to capture high-order connection relationships using GNNs. The other is to make fundamental innovations in the group representation paradigm, where CubeRec \cite{chen2022thinking} proposed a revolutionary idea of replacing point vectors with hypercubes (a subspace of preference intervals), which is more in line with the reality of achieving an "acceptable range" in group decision-making, providing a more powerful representation tool for understanding group preferences.
Although GNNs and advanced representation learning (such as hypercubes) enhance the model's ability to depict complex structures and preference intervals, most of these works still focus on "how to better aggregate the given individual preferences". \cite{wang2025unified} They fail to deeply reflect on the sociological nature of group decision-making, that is, they do not systematically incorporate social dynamic factors such as power structures, influence propagation, and role division within the group into the modeling framework. Their models are still based on the simplified assumption that "the group output is some function of the individual input".

Addressing these limitations, pioneering work proposed the Leadership-Dominant Preference Hypothesis \cite{gan2025large}, empirically validating leader groups where single members dominate decisions—providing a transformative advancement for group recommendation methodologies. 

Our approach improves the learning paradigm of group recommendation by using LLM-based agents to identify leadership groups.

\subsection{LLM-Agent for Recommendation Systems }

Recent LLMs breakthroughs in semantic reasoning and generation have revolutionized recommendation research. Key efforts focus on efficient LLMs \cite{bao2023tallrec} tuning for recommendation tasks while addressing deployment challenges like inference costs, hallucination, bias, and evaluation. This paradigm  \cite{shehmir2025llm4rec} enables natural interactions, open-intent understanding, cross-domain recommendations.  In recommendation systems, LLMs are driving paradigm shifts through three technical pathways:

LLM-as-Center \cite{bao2023tallrec, geng2022recommendation} employs LLMs as the core recommendation engine.

LLM-as-Enhancer \cite{zhang2025arise} augments traditional recommenders by generating semantic features. \cite{yang5603825patent} enhances the recommendation effect by integrating large language models into the knowledge graph.

Researchers \cite{hou2025generative} have proposed the construction of generative recommendation models by combining pre-trained generative models with user behavior or designing recommendation models within a generative framework. These models enable the system to receive and provide content that is closer to human-like, such as natural language, images, and so on.
LLM-as-Agent \cite{wang2025ruleagent} utilizes LLM-based agents as foundational components enabling natural interaction, explainability, and proactive collaboration. 

Through intelligent agents, large language models can participate more deeply in both of these paradigms. For instance, RecMind \cite{wang2024recmind} has developed a unified and fully-capable LLM intelligent agent that can directly generate recommendations based on the output of the LLM. AutoConcierge \cite{zeng2024automated} utilizes natural language conversations to understand user needs, collect user preferences, and ultimately provide explainable and personalized recommendations. UserSimulator \cite{yoon2024evaluating} has proposed an evaluation protocol, which assesses the effectiveness of LLM as a generative user simulator through five tasks, in order to measure to what extent these simulators can simulate real user behaviors.

Our approach utilizes LLM-based intelligent agents to fill the gaps in group recommendation identification. At the same time, it expands the application methods of intelligent agents in group recommendation.

% Recent LLMs breakthroughs in semantic reasoning and generation have revolutionized recommendation research. Key efforts focus on efficient LLMs \cite{bao2023tallrec} tuning for recommendation tasks while addressing deployment challenges like inference costs, hallucination, bias, and evaluation. This paradigm  \cite{shehmir2025llm4rec} enables natural interactions, open-intent understanding, cross-domain recommendations, and content generation, shifting systems from passive prediction to proactive perception and collaboration. 

\section{Conclusion}\label{}
% 本文提出DALI框架，首创融合大语言模型符号推理与神经表征学习的动态领导群组识别方法。其自主规则演化引擎与神经符号双流机制，精准建模复杂群组决策中的领导力动态。实证验证显示该框架在多种模型上实现稳定性能提升，基础方法改进尤为显著。闭环认知架构通过自优化持续精炼识别策略。未来将探索多类型群组动态建模与跨领域适配。 
% 本文提出DALI框架，首创大语言模型与神经网络的融合方法实现动态领导群组识别。其自主规则引擎与神经符号机制精准建模领导力动态。实验验证该框架在多模型上实现稳定性能提升，基础方法改进显著。闭环架构通过自优化持续精炼识别策略，未来将拓展至多类型群组动态建模与跨领域适配。 
This paper proposes DALI, a novel framework pioneering LLM-neural integration for dynamic leadership identification in group recommendation. Its autonomous rule engine and neuro-symbolic mechanism enable precise modeling of leadership dynamics. Experiments confirm consistent performance gains across models, with foundational approaches showing significant improvement. The closed-loop architecture continuously refines strategies through self-optimization. Future work extends to multi-type group dynamics and cross-domain adaptation. 

\bibliographystyle{cas-model2-names}

% Loading bibliography database
% \bibliography{cas-refs}

\begin{thebibliography}{66}
\expandafter\ifx\csname natexlab\endcsname\relax\def\natexlab#1{#1}\fi
\providecommand{\url}[1]{\texttt{#1}}
\providecommand{\href}[2]{#2}
\providecommand{\path}[1]{#1}
\providecommand{\DOIprefix}{doi:}
\providecommand{\ArXivprefix}{arXiv:}
\providecommand{\URLprefix}{URL: }
\providecommand{\Pubmedprefix}{pmid:}
\providecommand{\doi}[1]{\href{http://dx.doi.org/#1}{\path{#1}}}
\providecommand{\Pubmed}[1]{\href{pmid:#1}{\path{#1}}}
\providecommand{\bibinfo}[2]{#2}
\ifx\xfnm\relax \def\xfnm[#1]{\unskip,\space#1}\fi
%Type = Article
\bibitem[{Alvarado et~al.(2022)Alvarado, Htun, Jin and Verbert}]{alvarado2022systematic}
\bibinfo{author}{Alvarado, O.}, \bibinfo{author}{Htun, N.N.}, \bibinfo{author}{Jin, Y.}, \bibinfo{author}{Verbert, K.}, \bibinfo{year}{2022}.
\newblock \bibinfo{title}{A systematic review of interaction design strategies for group recommendation systems}.
\newblock \bibinfo{journal}{Proceedings of the ACM on Human-Computer Interaction} \bibinfo{volume}{6}, \bibinfo{pages}{1--51}.
%Type = Inproceedings
\bibitem[{Bao et~al.(2023)Bao, Zhang, Zhang, Wang, Feng and He}]{bao2023tallrec}
\bibinfo{author}{Bao, K.}, \bibinfo{author}{Zhang, J.}, \bibinfo{author}{Zhang, Y.}, \bibinfo{author}{Wang, W.}, \bibinfo{author}{Feng, F.}, \bibinfo{author}{He, X.}, \bibinfo{year}{2023}.
\newblock \bibinfo{title}{Tallrec: An effective and efficient tuning framework to align large language model with recommendation}, in: \bibinfo{booktitle}{Proceedings of the 17th ACM conference on recommender systems}, pp. \bibinfo{pages}{1007--1014}.
%Type = Article
\bibitem[{Blondel et~al.(2008)Blondel, Guillaume, Lambiotte and Lefebvre}]{Blondeletal2008}
\bibinfo{author}{Blondel, V.D.}, \bibinfo{author}{Guillaume, J.L.}, \bibinfo{author}{Lambiotte, R.}, \bibinfo{author}{Lefebvre, E.}, \bibinfo{year}{2008}.
\newblock \bibinfo{title}{Fast unfolding of communities in large networks}.
\newblock \bibinfo{journal}{J. Stat. Mech.-Theory Exp.} \bibinfo{volume}{2008}, \bibinfo{pages}{P10008}.
%Type = Inproceedings
\bibitem[{Cao et~al.(2018)Cao, He, Miao, An, Yang and Hong}]{cao2018attentive}
\bibinfo{author}{Cao, D.}, \bibinfo{author}{He, X.}, \bibinfo{author}{Miao, L.}, \bibinfo{author}{An, Y.}, \bibinfo{author}{Yang, C.}, \bibinfo{author}{Hong, R.}, \bibinfo{year}{2018}.
\newblock \bibinfo{title}{Attentive group recommendation}, in: \bibinfo{booktitle}{The 41st International ACM SIGIR conference on research \& development in information retrieval}, pp. \bibinfo{pages}{645--654}.
%Type = Article
\bibitem[{Cao et~al.(2019)Cao, He, Miao, Xiao, Chen and Xu}]{cao2019social}
\bibinfo{author}{Cao, D.}, \bibinfo{author}{He, X.}, \bibinfo{author}{Miao, L.}, \bibinfo{author}{Xiao, G.}, \bibinfo{author}{Chen, H.}, \bibinfo{author}{Xu, J.}, \bibinfo{year}{2019}.
\newblock \bibinfo{title}{Social-enhanced attentive group recommendation}.
\newblock \bibinfo{journal}{IEEE Transactions on Knowledge and Data Engineering} \bibinfo{volume}{33}, \bibinfo{pages}{1195--1209}.
%Type = Article
\bibitem[{Chen et~al.(2013)Chen, Wu and Fang}]{Chenetal2013}
\bibinfo{author}{Chen, Q.}, \bibinfo{author}{Wu, T.T.}, \bibinfo{author}{Fang, M.}, \bibinfo{year}{2013}.
\newblock \bibinfo{title}{Detecting local community structure in complex networks based on local degree central nodes}.
\newblock \bibinfo{journal}{Physica A.} \bibinfo{volume}{392}, \bibinfo{pages}{529--537}.
%Type = Inproceedings
\bibitem[{Chen et~al.(2022)Chen, Yin, Long, Nguyen, Wang and Wang}]{chen2022thinking}
\bibinfo{author}{Chen, T.}, \bibinfo{author}{Yin, H.}, \bibinfo{author}{Long, J.}, \bibinfo{author}{Nguyen, Q.V.H.}, \bibinfo{author}{Wang, Y.}, \bibinfo{author}{Wang, M.}, \bibinfo{year}{2022}.
\newblock \bibinfo{title}{Thinking inside the box: learning hypercube representations for group recommendation}, in: \bibinfo{booktitle}{Proceedings of the 45th International ACM SIGIR Conference on Research and Development in Information Retrieval}, pp. \bibinfo{pages}{1664--1673}.
%Type = Article
\bibitem[{Clauset et~al.(2004)Clauset, Newman and Moore}]{Clausetetal2004}
\bibinfo{author}{Clauset, A.}, \bibinfo{author}{Newman, M.E.J.}, \bibinfo{author}{Moore, C.}, \bibinfo{year}{2004}.
\newblock \bibinfo{title}{Finding community structure in very large networks}.
\newblock \bibinfo{journal}{Phys. Rev. E.} \bibinfo{volume}{70}, \bibinfo{pages}{066111}.
%Type = Article
\bibitem[{Danon et~al.(2005)Danon, Diaz-Guilera, Duch and Arenas}]{Danonetal2005}
\bibinfo{author}{Danon, L.}, \bibinfo{author}{Diaz-Guilera, A.}, \bibinfo{author}{Duch, J.}, \bibinfo{author}{Arenas, A.}, \bibinfo{year}{2005}.
\newblock \bibinfo{title}{Comparing community structure identification}.
\newblock \bibinfo{journal}{J. Stat. Mech.-Theory Exp.} , \bibinfo{pages}{P09008}.
%Type = Inproceedings
\bibitem[{Deng et~al.(2021)Deng, Li, Liu, Ali and Shao}]{deng2021knowledge}
\bibinfo{author}{Deng, Z.}, \bibinfo{author}{Li, C.}, \bibinfo{author}{Liu, S.}, \bibinfo{author}{Ali, W.}, \bibinfo{author}{Shao, J.}, \bibinfo{year}{2021}.
\newblock \bibinfo{title}{Knowledge-aware group representation learning for group recommendation}, in: \bibinfo{booktitle}{2021 IEEE 37th International Conference on Data Engineering (ICDE)}, \bibinfo{organization}{IEEE}. pp. \bibinfo{pages}{1571--1582}.
%Type = Article
\bibitem[{Fabio et~al.(2013)Fabio, Fabio and Carlo}]{Fabioetal2013}
\bibinfo{author}{Fabio, D.R.}, \bibinfo{author}{Fabio, D.}, \bibinfo{author}{Carlo, P.}, \bibinfo{year}{2013}.
\newblock \bibinfo{title}{Profiling core-periphery network structure by random walkers}.
\newblock \bibinfo{journal}{Sci. Rep.} \bibinfo{volume}{3}, \bibinfo{pages}{1467}.
%Type = Article
\bibitem[{Fabricio and Liang(2013)}]{FabricioLiang2013}
\bibinfo{author}{Fabricio, B.}, \bibinfo{author}{Liang, Z.}, \bibinfo{year}{2013}.
\newblock \bibinfo{title}{Fuzzy community structure detection by particle competition and cooperation}.
\newblock \bibinfo{journal}{Soft Comput.} \bibinfo{volume}{17}, \bibinfo{pages}{659--673}.
%Type = Article
\bibitem[{Fortunato(2010)}]{Fortunato2010}
\bibinfo{author}{Fortunato, S.}, \bibinfo{year}{2010}.
\newblock \bibinfo{title}{Community detection in graphs}.
\newblock \bibinfo{journal}{Phys. Rep.-Rev. Sec. Phys. Lett.} \bibinfo{volume}{486}, \bibinfo{pages}{75--174}.
%Type = Article
\bibitem[{Fortunato and Barthelemy(2007)}]{FortunatoBarthelemy2007}
\bibinfo{author}{Fortunato, S.}, \bibinfo{author}{Barthelemy, M.}, \bibinfo{year}{2007}.
\newblock \bibinfo{title}{Resolution limit in community detection}.
\newblock \bibinfo{journal}{Proc. Natl. Acad. Sci. U. S. A.} \bibinfo{volume}{104}, \bibinfo{pages}{36--41}.
%Type = Article
\bibitem[{Gan et~al.(2025)Gan, Gao, Li, Wang, Guo, Jiang and Song}]{gan2025large}
\bibinfo{author}{Gan, D.}, \bibinfo{author}{Gao, M.}, \bibinfo{author}{Li, W.}, \bibinfo{author}{Wang, Z.}, \bibinfo{author}{Guo, L.}, \bibinfo{author}{Jiang, F.}, \bibinfo{author}{Song, Y.}, \bibinfo{year}{2025}.
\newblock \bibinfo{title}{Large: A leadership perception framework for group recommendation}.
\newblock \bibinfo{journal}{Expert Systems with Applications} \bibinfo{volume}{260}, \bibinfo{pages}{125416}.
%Type = Article
\bibitem[{Gao et~al.(2025)Gao, Yu, Chen, Ye, Zhang and Yin}]{gao2025graph}
\bibinfo{author}{Gao, X.}, \bibinfo{author}{Yu, J.}, \bibinfo{author}{Chen, T.}, \bibinfo{author}{Ye, G.}, \bibinfo{author}{Zhang, W.}, \bibinfo{author}{Yin, H.}, \bibinfo{year}{2025}.
\newblock \bibinfo{title}{Graph condensation: A survey}.
\newblock \bibinfo{journal}{IEEE Transactions on Knowledge and Data Engineering} \bibinfo{volume}{37}, \bibinfo{pages}{1819--1837}.
%Type = Inproceedings
\bibitem[{Geng et~al.(2022)Geng, Liu, Fu, Ge and Zhang}]{geng2022recommendation}
\bibinfo{author}{Geng, S.}, \bibinfo{author}{Liu, S.}, \bibinfo{author}{Fu, Z.}, \bibinfo{author}{Ge, Y.}, \bibinfo{author}{Zhang, Y.}, \bibinfo{year}{2022}.
\newblock \bibinfo{title}{Recommendation as language processing (rlp): A unified pretrain, personalized prompt \& predict paradigm (p5)}, in: \bibinfo{booktitle}{Proceedings of the 16th ACM conference on recommender systems}, pp. \bibinfo{pages}{299--315}.
%Type = Article
\bibitem[{Gregory(2011)}]{Gregory2011}
\bibinfo{author}{Gregory, S.}, \bibinfo{year}{2011}.
\newblock \bibinfo{title}{Fuzzy overlapping communities in networks}.
\newblock \bibinfo{journal}{J. Stat. Mech.-Theory Exp.} , \bibinfo{pages}{P02017}.
%Type = Article
\bibitem[{Havens et~al.(2013)Havens, Bezdek, Leckie, Ramamohanarao and Palaniswami}]{Havensetal2013}
\bibinfo{author}{Havens, T.C.}, \bibinfo{author}{Bezdek, J.C.}, \bibinfo{author}{Leckie, C.}, \bibinfo{author}{Ramamohanarao, K.}, \bibinfo{author}{Palaniswami, M.}, \bibinfo{year}{2013}.
\newblock \bibinfo{title}{A soft modularity function for detecting fuzzy communities in social networks}.
\newblock \bibinfo{journal}{IEEE Trans. Fuzzy Syst.} \bibinfo{volume}{21}, \bibinfo{pages}{1170--1175}.
%Type = Inproceedings
\bibitem[{Hou et~al.(2025)Hou, Zhang, Sheng, Yang, Wang, Chua and McAuley}]{hou2025generative}
\bibinfo{author}{Hou, Y.}, \bibinfo{author}{Zhang, A.}, \bibinfo{author}{Sheng, L.}, \bibinfo{author}{Yang, Z.}, \bibinfo{author}{Wang, X.}, \bibinfo{author}{Chua, T.S.}, \bibinfo{author}{McAuley, J.}, \bibinfo{year}{2025}.
\newblock \bibinfo{title}{Generative recommendation models: Progress and directions}, in: \bibinfo{booktitle}{Companion Proceedings of the ACM on Web Conference 2025}, pp. \bibinfo{pages}{13--16}.
%Type = Article
\bibitem[{Huang et~al.(2020)Huang, Xu, Zhu and Zhou}]{huang2020efficient}
\bibinfo{author}{Huang, Z.}, \bibinfo{author}{Xu, X.}, \bibinfo{author}{Zhu, H.}, \bibinfo{author}{Zhou, M.}, \bibinfo{year}{2020}.
\newblock \bibinfo{title}{An efficient group recommendation model with multiattention-based neural networks}.
\newblock \bibinfo{journal}{IEEE Transactions on Neural Networks and Learning Systems} \bibinfo{volume}{31}, \bibinfo{pages}{4461--4474}.
%Type = Inproceedings
\bibitem[{Hullermeier and Rifqi(2009)}]{HullermeierRifqi2009}
\bibinfo{author}{Hullermeier, E.}, \bibinfo{author}{Rifqi, M.}, \bibinfo{year}{2009}.
\newblock \bibinfo{title}{A fuzzy variant of the rand index for comparing clustering structures}, in: \bibinfo{booktitle}{in Proc. IFSA/EUSFLAT Conf.}, pp. \bibinfo{pages}{1294--1298}.
%Type = Inproceedings
\bibitem[{Jia et~al.(2021)Jia, Zhou, Dong and Pan}]{jia2021hypergraph}
\bibinfo{author}{Jia, R.}, \bibinfo{author}{Zhou, X.}, \bibinfo{author}{Dong, L.}, \bibinfo{author}{Pan, S.}, \bibinfo{year}{2021}.
\newblock \bibinfo{title}{Hypergraph convolutional network for group recommendation}, in: \bibinfo{booktitle}{2021 ieee international conference on data mining (icdm)}, \bibinfo{organization}{IEEE}. pp. \bibinfo{pages}{260--269}.
%Type = Article
\bibitem[{Krishnamoorthi and Shyam(2026)}]{krishnamoorthi2026self}
\bibinfo{author}{Krishnamoorthi, S.}, \bibinfo{author}{Shyam, G.K.}, \bibinfo{year}{2026}.
\newblock \bibinfo{title}{Self-supervised social attentive deep reinforcement learning-based group recommender system}.
\newblock \bibinfo{journal}{Engineering Applications of Artificial Intelligence} \bibinfo{volume}{165}, \bibinfo{pages}{113053}.
%Type = Article
\bibitem[{Lancichinetti and Fortunato(2009)}]{LancichinettiFortunato2009}
\bibinfo{author}{Lancichinetti, A.}, \bibinfo{author}{Fortunato, S.}, \bibinfo{year}{2009}.
\newblock \bibinfo{title}{Benchmarks for testing community detection algorithms on directed and weighted graphs with overlapping communities}.
\newblock \bibinfo{journal}{Phys. Rev. E.} \bibinfo{volume}{80}, \bibinfo{pages}{016118}.
%Type = Article
\bibitem[{Lancichinetti et~al.(2008)Lancichinetti, Fortunato and Radicchi}]{Lancichinettietal2008}
\bibinfo{author}{Lancichinetti, A.}, \bibinfo{author}{Fortunato, S.}, \bibinfo{author}{Radicchi, F.}, \bibinfo{year}{2008}.
\newblock \bibinfo{title}{Benchmark graphs for testing community detection algorithms}.
\newblock \bibinfo{journal}{Phys. Rev. E.} \bibinfo{volume}{78}, \bibinfo{pages}{046110}.
%Type = Article
\bibitem[{Li et~al.(2013)Li, Wang and Eustace}]{Lietal2013}
\bibinfo{author}{Li, J.}, \bibinfo{author}{Wang, X.}, \bibinfo{author}{Eustace, J.}, \bibinfo{year}{2013}.
\newblock \bibinfo{title}{Detecting overlapping communities by seed community in weighted complex networks}.
\newblock \bibinfo{journal}{Physica A.} \bibinfo{volume}{392}, \bibinfo{pages}{6125--6134}.
%Type = Inproceedings
\bibitem[{Li et~al.(2023)Li, Wang, Lai and Yuan}]{li2023self}
\bibinfo{author}{Li, K.}, \bibinfo{author}{Wang, C.D.}, \bibinfo{author}{Lai, J.H.}, \bibinfo{author}{Yuan, H.}, \bibinfo{year}{2023}.
\newblock \bibinfo{title}{Self-supervised group graph collaborative filtering for group recommendation}, in: \bibinfo{booktitle}{Proceedings of the sixteenth ACM international conference on web search and data mining}, pp. \bibinfo{pages}{69--77}.
%Type = Article
\bibitem[{Li et~al.(2024)Li, Liu, Satapathy, Wang and Cambria}]{li2024recent}
\bibinfo{author}{Li, Y.}, \bibinfo{author}{Liu, K.}, \bibinfo{author}{Satapathy, R.}, \bibinfo{author}{Wang, S.}, \bibinfo{author}{Cambria, E.}, \bibinfo{year}{2024}.
\newblock \bibinfo{title}{Recent developments in recommender systems: A survey}.
\newblock \bibinfo{journal}{IEEE Computational Intelligence Magazine} \bibinfo{volume}{19}, \bibinfo{pages}{78--95}.
%Type = Article
\bibitem[{Liu(2010)}]{Liu2010}
\bibinfo{author}{Liu, J.}, \bibinfo{year}{2010}.
\newblock \bibinfo{title}{Fuzzy modularity and fuzzy community structure in networks}.
\newblock \bibinfo{journal}{Eur. Phys. J. B.} \bibinfo{volume}{77}, \bibinfo{pages}{547--557}.
%Type = Article
\bibitem[{Liu et~al.(2014)Liu, Pellegrini and Wang}]{Liuetal2014}
\bibinfo{author}{Liu, W.}, \bibinfo{author}{Pellegrini, M.}, \bibinfo{author}{Wang, X.}, \bibinfo{year}{2014}.
\newblock \bibinfo{title}{Detecting communities based on network topology}.
\newblock \bibinfo{journal}{Sci. Rep.} \bibinfo{volume}{4}, \bibinfo{pages}{5739}.
%Type = Article
\bibitem[{Lou et~al.(2013)Lou, Li and Zhao}]{Louetal2013}
\bibinfo{author}{Lou, H.}, \bibinfo{author}{Li, S.}, \bibinfo{author}{Zhao, Y.}, \bibinfo{year}{2013}.
\newblock \bibinfo{title}{Detecting community structure using label propagation with weighted coherent neighborhood propinquity}.
\newblock \bibinfo{journal}{Physica A.} \bibinfo{volume}{392}, \bibinfo{pages}{3095--3105}.
%Type = Inproceedings
\bibitem[{Malecek and Peska(2021)}]{malecek2021fairness}
\bibinfo{author}{Malecek, L.}, \bibinfo{author}{Peska, L.}, \bibinfo{year}{2021}.
\newblock \bibinfo{title}{Fairness-preserving group recommendations with user weighting}, in: \bibinfo{booktitle}{Adjunct Proceedings of the 29th ACM Conference on User Modeling, Adaptation and Personalization}, pp. \bibinfo{pages}{4--9}.
%Type = Article
\bibitem[{Nepusz et~al.(2008)Nepusz, Petr\'oczi, N\'egyessy and Bazs\'o}]{Nepuszetal2008}
\bibinfo{author}{Nepusz, T.}, \bibinfo{author}{Petr\'oczi, A.}, \bibinfo{author}{N\'egyessy, L.}, \bibinfo{author}{Bazs\'o, F.}, \bibinfo{year}{2008}.
\newblock \bibinfo{title}{Fuzzy communities and the concept of bridgeness in complex networks}.
\newblock \bibinfo{journal}{Phys. Rev. E.} \bibinfo{volume}{77}, \bibinfo{pages}{016107}.
%Type = Misc
\bibitem[{Newman(2013)}]{Newman2013}
\bibinfo{author}{Newman, M.E.J.}, \bibinfo{year}{2013}.
\newblock \bibinfo{title}{Network data}.
\newblock \bibinfo{howpublished}{\url{http://www-personal.umich.edu/~mejn/netdata/}}.
%Type = Article
\bibitem[{Newman and Girvan(2004)}]{NewmanGirvan2004}
\bibinfo{author}{Newman, M.E.J.}, \bibinfo{author}{Girvan, M.}, \bibinfo{year}{2004}.
\newblock \bibinfo{title}{Finding and evaluating community structure in networks}.
\newblock \bibinfo{journal}{Phys. Rev. E.} \bibinfo{volume}{69}, \bibinfo{pages}{026113}.
%Type = Article
\bibitem[{Nguyen et~al.(2025)Nguyen, Huynh, Ren, Nguyen, Liew, Yin and Nguyen}]{nguyen2025survey}
\bibinfo{author}{Nguyen, T.T.}, \bibinfo{author}{Huynh, T.T.}, \bibinfo{author}{Ren, Z.}, \bibinfo{author}{Nguyen, P.L.}, \bibinfo{author}{Liew, A.W.C.}, \bibinfo{author}{Yin, H.}, \bibinfo{author}{Nguyen, Q.V.H.}, \bibinfo{year}{2025}.
\newblock \bibinfo{title}{A survey of machine unlearning}.
\newblock \bibinfo{journal}{ACM Transactions on Intelligent Systems and Technology} \bibinfo{volume}{16}, \bibinfo{pages}{1--46}.
%Type = Article
\bibitem[{Psorakis et~al.(2011)Psorakis, Roberts, Ebden and Sheldon}]{Psorakisetal2011}
\bibinfo{author}{Psorakis, I.}, \bibinfo{author}{Roberts, S.}, \bibinfo{author}{Ebden, M.}, \bibinfo{author}{Sheldon, B.}, \bibinfo{year}{2011}.
\newblock \bibinfo{title}{Overlapping community detection using bayesian non-negative matrix factorization}.
\newblock \bibinfo{journal}{Phys. Rev. E.} \bibinfo{volume}{83}, \bibinfo{pages}{066114}.
%Type = Article
\bibitem[{Raghavan et~al.(2007)Raghavan, Albert and Kumara}]{Raghavanetal2007}
\bibinfo{author}{Raghavan, U.}, \bibinfo{author}{Albert, R.}, \bibinfo{author}{Kumara, S.}, \bibinfo{year}{2007}.
\newblock \bibinfo{title}{Near linear time algorithm to detect community structures in large-scale networks}.
\newblock \bibinfo{journal}{Phys. Rev E.} \bibinfo{volume}{76}, \bibinfo{pages}{036106}.
%Type = Inproceedings
\bibitem[{Sankar et~al.(2020)Sankar, Wu, Wu, Zhang, Yang and Sundaram}]{sankar2020groupim}
\bibinfo{author}{Sankar, A.}, \bibinfo{author}{Wu, Y.}, \bibinfo{author}{Wu, Y.}, \bibinfo{author}{Zhang, W.}, \bibinfo{author}{Yang, H.}, \bibinfo{author}{Sundaram, H.}, \bibinfo{year}{2020}.
\newblock \bibinfo{title}{Groupim: A mutual information maximization framework for neural group recommendation}, in: \bibinfo{booktitle}{Proceedings of the 43rd International ACM SIGIR conference on research and development in Information Retrieval}, pp. \bibinfo{pages}{1279--1288}.
%Type = Inproceedings
\bibitem[{Sato(2022)}]{sato2022enumerating}
\bibinfo{author}{Sato, R.}, \bibinfo{year}{2022}.
\newblock \bibinfo{title}{Enumerating fair packages for group recommendations}, in: \bibinfo{booktitle}{Proceedings of the Fifteenth ACM International Conference on Web Search and Data Mining}, pp. \bibinfo{pages}{870--878}.
%Type = Article
\bibitem[{Shehmir and Kashef(2025)}]{shehmir2025llm4rec}
\bibinfo{author}{Shehmir, S.}, \bibinfo{author}{Kashef, R.}, \bibinfo{year}{2025}.
\newblock \bibinfo{title}{Llm4rec: A comprehensive survey on the integration of large language models in recommender systems—approaches, applications and challenges}.
\newblock \bibinfo{journal}{Future Internet} \bibinfo{volume}{17}, \bibinfo{pages}{252}.
%Type = Article
\bibitem[{Sobolevsky and Campari(2014)}]{SobolevskyCampari2014}
\bibinfo{author}{Sobolevsky, S.}, \bibinfo{author}{Campari, R.}, \bibinfo{year}{2014}.
\newblock \bibinfo{title}{General optimization technique for high-quality community detection in complex networks}.
\newblock \bibinfo{journal}{Phys. Rev. E.} \bibinfo{volume}{90}, \bibinfo{pages}{012811}.
%Type = Article
\bibitem[{Stratigi et~al.(2023)Stratigi, Pitoura and Stefanidis}]{stratigi2023squirrel}
\bibinfo{author}{Stratigi, M.}, \bibinfo{author}{Pitoura, E.}, \bibinfo{author}{Stefanidis, K.}, \bibinfo{year}{2023}.
\newblock \bibinfo{title}{Squirrel: A framework for sequential group recommendations through reinforcement learning}.
\newblock \bibinfo{journal}{Information Systems} \bibinfo{volume}{112}, \bibinfo{pages}{102128}.
%Type = Article
\bibitem[{Sun et~al.(2011)Sun, Gao and Han}]{Sunetal2011}
\bibinfo{author}{Sun, P.}, \bibinfo{author}{Gao, L.}, \bibinfo{author}{Han, S.}, \bibinfo{year}{2011}.
\newblock \bibinfo{title}{Identification of overlapping and non-overlapping community structure by fuzzy clustering in complex networks}.
\newblock \bibinfo{journal}{Inf. Sci.} \bibinfo{volume}{181}, \bibinfo{pages}{1060--1071}.
%Type = Article
\bibitem[{Vehlow et~al.(2013)Vehlow, Reinhardt and Weiskopf}]{Vehlowetal2013}
\bibinfo{author}{Vehlow, C.}, \bibinfo{author}{Reinhardt, T.}, \bibinfo{author}{Weiskopf, D.}, \bibinfo{year}{2013}.
\newblock \bibinfo{title}{Visualizing fuzzy overlapping communities in networks}.
\newblock \bibinfo{journal}{IEEE Trans. Vis. Comput. Graph.} \bibinfo{volume}{19}, \bibinfo{pages}{2486--2495}.
%Type = Inproceedings
\bibitem[{Vinh~Tran et~al.(2019)Vinh~Tran, Nguyen~Pham, Tay, Liu, Cong and Li}]{vinh2019interact}
\bibinfo{author}{Vinh~Tran, L.}, \bibinfo{author}{Nguyen~Pham, T.A.}, \bibinfo{author}{Tay, Y.}, \bibinfo{author}{Liu, Y.}, \bibinfo{author}{Cong, G.}, \bibinfo{author}{Li, X.}, \bibinfo{year}{2019}.
\newblock \bibinfo{title}{Interact and decide: Medley of sub-attention networks for effective group recommendation}, in: \bibinfo{booktitle}{Proceedings of the 42nd International ACM SIGIR conference on research and development in information retrieval}, pp. \bibinfo{pages}{255--264}.
%Type = Article
\bibitem[{\v{S}ubelj and Bajec(2011a)}]{SubeljBajec2011a}
\bibinfo{author}{\v{S}ubelj, L.}, \bibinfo{author}{Bajec, M.}, \bibinfo{year}{2011}a.
\newblock \bibinfo{title}{Robust network community detection using balanced propagation}.
\newblock \bibinfo{journal}{Eur. Phys. J. B.} \bibinfo{volume}{81}, \bibinfo{pages}{353--362}.
%Type = Article
\bibitem[{\v{S}ubelj and Bajec(2011b)}]{SubeljBajec2011b}
\bibinfo{author}{\v{S}ubelj, L.}, \bibinfo{author}{Bajec, M.}, \bibinfo{year}{2011}b.
\newblock \bibinfo{title}{Unfolding communities in large complex networks: Combining defensive and offensive label propagation for core extraction}.
\newblock \bibinfo{journal}{Phys. Rev. E.} \bibinfo{volume}{83}, \bibinfo{pages}{036103}.
%Type = Article
\bibitem[{\v{S}ubelj and Bajec(2012)}]{SubeljBajec2012}
\bibinfo{author}{\v{S}ubelj, L.}, \bibinfo{author}{Bajec, M.}, \bibinfo{year}{2012}.
\newblock \bibinfo{title}{Ubiquitousness of link-density and link-pattern communities in real-world networks}.
\newblock \bibinfo{journal}{Eur. Phys. J. B.} \bibinfo{volume}{85}, \bibinfo{pages}{1--11}.
%Type = Article
\bibitem[{Wang et~al.(2013)Wang, Liu, Liu and Pan}]{Wangetal2013}
\bibinfo{author}{Wang, W.}, \bibinfo{author}{Liu, D.}, \bibinfo{author}{Liu, X.}, \bibinfo{author}{Pan, L.}, \bibinfo{year}{2013}.
\newblock \bibinfo{title}{Fuzzy overlapping community detection based on local random walk and multidimensional scaling}.
\newblock \bibinfo{journal}{Physica A.} \bibinfo{volume}{392}, \bibinfo{pages}{6578--6586}.
%Type = Article
\bibitem[{Wang et~al.(2025a)Wang, Lai, Liu, Wang and Gao}]{wang2025unified}
\bibinfo{author}{Wang, X.}, \bibinfo{author}{Lai, N.}, \bibinfo{author}{Liu, P.}, \bibinfo{author}{Wang, Z.}, \bibinfo{author}{Gao, M.}, \bibinfo{year}{2025}a.
\newblock \bibinfo{title}{A unified adaptive graph structure generation method for spatio-temporal graph forecasting}.
\newblock \bibinfo{journal}{Knowledge-Based Systems} \bibinfo{volume}{309}, \bibinfo{pages}{112811}.
%Type = Article
\bibitem[{Wang and Li(2013)}]{WangLi2013}
\bibinfo{author}{Wang, X.}, \bibinfo{author}{Li, J.}, \bibinfo{year}{2013}.
\newblock \bibinfo{title}{Detecting communities by the core-vertex and intimate degree in complex networks}.
\newblock \bibinfo{journal}{Physica A.} \bibinfo{volume}{392}, \bibinfo{pages}{2555--2563}.
%Type = Inproceedings
\bibitem[{Wang et~al.(2024)Wang, Jiang, Chen, Yang, Zhou, Cho, Fan, Lu, Huang and Yang}]{wang2024recmind}
\bibinfo{author}{Wang, Y.}, \bibinfo{author}{Jiang, Z.}, \bibinfo{author}{Chen, Z.}, \bibinfo{author}{Yang, F.}, \bibinfo{author}{Zhou, Y.}, \bibinfo{author}{Cho, E.}, \bibinfo{author}{Fan, X.}, \bibinfo{author}{Lu, Y.}, \bibinfo{author}{Huang, X.}, \bibinfo{author}{Yang, Y.}, \bibinfo{year}{2024}.
\newblock \bibinfo{title}{Recmind: Large language model powered agent for recommendation}, in: \bibinfo{booktitle}{Findings of the Association for Computational Linguistics: NAACL 2024}, pp. \bibinfo{pages}{4351--4364}.
%Type = Article
\bibitem[{Wang et~al.(2025b)Wang, Gao, Yu, Hou, Sadiq and Yin}]{wang2025ruleagent}
\bibinfo{author}{Wang, Z.}, \bibinfo{author}{Gao, M.}, \bibinfo{author}{Yu, J.}, \bibinfo{author}{Hou, Y.}, \bibinfo{author}{Sadiq, S.}, \bibinfo{author}{Yin, H.}, \bibinfo{year}{2025}b.
\newblock \bibinfo{title}{Ruleagent: Discovering rules for recommendation denoising with autonomous language agents}.
\newblock \bibinfo{journal}{arXiv preprint arXiv:2503.23374} .
%Type = Article
\bibitem[{Wang et~al.(2026)Wang, Gao, Yu, Sadiq, Yin and Liu}]{wang2026graph}
\bibinfo{author}{Wang, Z.}, \bibinfo{author}{Gao, M.}, \bibinfo{author}{Yu, J.}, \bibinfo{author}{Sadiq, S.}, \bibinfo{author}{Yin, H.}, \bibinfo{author}{Liu, L.}, \bibinfo{year}{2026}.
\newblock \bibinfo{title}{When graph contrastive learning backfires: Spectral vulnerability and defense in recommendation}.
\newblock \bibinfo{journal}{ACM Transactions on Information Systems} \bibinfo{volume}{44}, \bibinfo{pages}{1--30}.
%Type = Inproceedings
\bibitem[{Wu et~al.(2023)Wu, Xiong, Zhang, Jiao, Zhang, Zhu and Yu}]{wu2023consrec}
\bibinfo{author}{Wu, X.}, \bibinfo{author}{Xiong, Y.}, \bibinfo{author}{Zhang, Y.}, \bibinfo{author}{Jiao, Y.}, \bibinfo{author}{Zhang, J.}, \bibinfo{author}{Zhu, Y.}, \bibinfo{author}{Yu, P.S.}, \bibinfo{year}{2023}.
\newblock \bibinfo{title}{Consrec: Learning consensus behind interactions for group recommendation}, in: \bibinfo{booktitle}{Proceedings of the acm web conference 2023}, pp. \bibinfo{pages}{240--250}.
%Type = Article
\bibitem[{Yang et~al.(2026)Yang, Wu and Wen}]{yang5603825patent}
\bibinfo{author}{Yang, P.}, \bibinfo{author}{Wu, X.}, \bibinfo{author}{Wen, P.}, \bibinfo{year}{2026}.
\newblock \bibinfo{title}{Patent technology knowledge recommendation by integrating large language models and knowledge graphs}.
\newblock \bibinfo{journal}{Available at SSRN 5603825} .
%Type = Article
\bibitem[{Yang et~al.(2024)Yang, Wang, Huang, Guo, Shi, Han, Feng, Song, Liang and Tang}]{yang2024optibench}
\bibinfo{author}{Yang, Z.}, \bibinfo{author}{Wang, Y.}, \bibinfo{author}{Huang, Y.}, \bibinfo{author}{Guo, Z.}, \bibinfo{author}{Shi, W.}, \bibinfo{author}{Han, X.}, \bibinfo{author}{Feng, L.}, \bibinfo{author}{Song, L.}, \bibinfo{author}{Liang, X.}, \bibinfo{author}{Tang, J.}, \bibinfo{year}{2024}.
\newblock \bibinfo{title}{Optibench meets resocratic: Measure and improve llms for optimization modeling}.
\newblock \bibinfo{journal}{arXiv preprint arXiv:2407.09887} .
%Type = Inproceedings
\bibitem[{Ye et~al.(2025)Ye, Wu, Wang, Chen, Zheng and He}]{ye2025disentangled}
\bibinfo{author}{Ye, G.}, \bibinfo{author}{Wu, W.}, \bibinfo{author}{Wang, G.}, \bibinfo{author}{Chen, X.}, \bibinfo{author}{Zheng, H.}, \bibinfo{author}{He, L.}, \bibinfo{year}{2025}.
\newblock \bibinfo{title}{Disentangled modeling of preferences and social influence for group recommendation}, in: \bibinfo{booktitle}{Proceedings of the AAAI Conference on Artificial Intelligence}, pp. \bibinfo{pages}{13052--13060}.
%Type = Inproceedings
\bibitem[{Yin et~al.(2023)Yin, Wang, Zhang, Meng, Yang, Lu and Luo}]{yin2023beyond}
\bibinfo{author}{Yin, G.}, \bibinfo{author}{Wang, X.}, \bibinfo{author}{Zhang, H.}, \bibinfo{author}{Meng, C.}, \bibinfo{author}{Yang, Y.}, \bibinfo{author}{Lu, K.}, \bibinfo{author}{Luo, Y.}, \bibinfo{year}{2023}.
\newblock \bibinfo{title}{Beyond individuals: Modeling mutual and multiple interactions for inductive link prediction between groups}, in: \bibinfo{booktitle}{Proceedings of the Sixteenth ACM International Conference on Web Search and Data Mining}, pp. \bibinfo{pages}{751--759}.
%Type = Inproceedings
\bibitem[{Yoon et~al.(2024)Yoon, He, Echterhoff and McAuley}]{yoon2024evaluating}
\bibinfo{author}{Yoon, S.e.}, \bibinfo{author}{He, Z.}, \bibinfo{author}{Echterhoff, J.}, \bibinfo{author}{McAuley, J.}, \bibinfo{year}{2024}.
\newblock \bibinfo{title}{Evaluating large language models as generative user simulators for conversational recommendation}, in: \bibinfo{booktitle}{Proceedings of the 2024 Conference of the North American Chapter of the Association for Computational Linguistics: Human Language Technologies (Volume 1: Long Papers)}, pp. \bibinfo{pages}{1490--1504}.
%Type = Inproceedings
\bibitem[{Zeng et~al.(2024)Zeng, Rajasekharan, Padalkar, Basu, Arias and Gupta}]{zeng2024automated}
\bibinfo{author}{Zeng, Y.}, \bibinfo{author}{Rajasekharan, A.}, \bibinfo{author}{Padalkar, P.}, \bibinfo{author}{Basu, K.}, \bibinfo{author}{Arias, J.}, \bibinfo{author}{Gupta, G.}, \bibinfo{year}{2024}.
\newblock \bibinfo{title}{Automated interactive domain-specific conversational agents that understand human dialogs}, in: \bibinfo{booktitle}{International Symposium on Practical Aspects of Declarative Languages}, \bibinfo{organization}{Springer}. pp. \bibinfo{pages}{204--222}.
%Type = Article
\bibitem[{Zhang et~al.(2007)Zhang, Wang and Zhang}]{Zhangetal2007}
\bibinfo{author}{Zhang, S.}, \bibinfo{author}{Wang, R.}, \bibinfo{author}{Zhang, X.}, \bibinfo{year}{2007}.
\newblock \bibinfo{title}{Identification of overlapping community structure in complex networks using fuzzy c-means clustering}.
\newblock \bibinfo{journal}{Physica A.} \bibinfo{volume}{374}, \bibinfo{pages}{483--490}.
%Type = Article
\bibitem[{Zhang et~al.(2025)Zhang, Wang, Chen, Wang, Zeng, Lin, Han, Sun and Lu}]{zhang2025arise}
\bibinfo{author}{Zhang, Y.}, \bibinfo{author}{Wang, T.}, \bibinfo{author}{Chen, S.}, \bibinfo{author}{Wang, K.}, \bibinfo{author}{Zeng, X.}, \bibinfo{author}{Lin, H.}, \bibinfo{author}{Han, X.}, \bibinfo{author}{Sun, L.}, \bibinfo{author}{Lu, C.}, \bibinfo{year}{2025}.
\newblock \bibinfo{title}{Arise: Towards knowledge-augmented reasoning via risk-adaptive search}.
\newblock \bibinfo{journal}{arXiv preprint arXiv:2504.10893} .
%Type = Inproceedings
\bibitem[{Zhang and Yeung(2012)}]{ZhangYeung2012}
\bibinfo{author}{Zhang, Y.}, \bibinfo{author}{Yeung, D.}, \bibinfo{year}{2012}.
\newblock \bibinfo{title}{Overlapping community detection via bounded nonnegative matrix tri-factorization}, in: \bibinfo{booktitle}{In Proc. ACM SIGKDD Conf.}, pp. \bibinfo{pages}{606--614}.

\end{thebibliography}

% Biography
%\bio{}
% Here goes the biography details.
%\endbio

%\bio{pic1}
% Here goes the biography details.
%\endbio

\end{document}